\documentclass[iop,ap]{emulateapj}

\usepackage{natbib}
\usepackage{amsmath,amssymb}
\usepackage{apjfonts}
\usepackage{xspace}
\usepackage{rotating}
\usepackage[usenames]{color}
\usepackage{color}

\usepackage[plainpages=false, colorlinks=true, anchorcolor=blue, linkcolor=blue, citecolor=blue, bookmarks=false]{hyperref}
\shorttitle{NANOGrav Nine-year Isotropic GWB Limit}

\newcommand{\fbend}{f_\mathrm{bend}}
\newcommand{\Msun}{M_\odot}

\newcommand{\psrchive}{{\sc psrchive}\xspace}

\def\be{\begin{equation}}
\def\ee{\end{equation}}
\newcommand{\bb}{\begin{bmatrix}}
\newcommand{\eb}{\end{bmatrix}}
\newcommand{\lp}{\left(}
\newcommand{\rp}{\right)}
\newcommand{\msJ}{\mathcal{J}}
\newcommand{\msun}{M_{\odot}}

\def\bea{\begin{eqnarray}}
\def\eea{\end{eqnarray}}

\defcitealias{mop14}{MOP14}
\defcitealias{rws+14}{RWS14}
\defcitealias{s13}{S13}
\defcitealias{dfg+13}{DFG13}
\defcitealias{ltm+15}{LTM15}

\begin{document}

\title{The NANOGrav Nine-year Data Set: \\ Limits on the Isotropic 
Stochastic Gravitational Wave Background}
 
\author{
Z.~Arzoumanian\altaffilmark{1}, 
A.~Brazier\altaffilmark{2}, 
S.~Burke-Spolaor\altaffilmark{3,27},
S.~J.~Chamberlin\altaffilmark{4},
S.~Chatterjee\altaffilmark{2},
B.~Christy\altaffilmark{5},
J.~M.~Cordes\altaffilmark{2},
N.~J.~Cornish\altaffilmark{6}, 
P.~B.~Demorest\altaffilmark{3},
X.~Deng\altaffilmark{7},
T.~Dolch\altaffilmark{2, 25},
J.~A.~Ellis\altaffilmark{8,28},
R.~D.~Ferdman\altaffilmark{9},
E.~Fonseca\altaffilmark{10},
N.~Garver-Daniels\altaffilmark{11},
F.~Jenet\altaffilmark{12}, 
G.~Jones\altaffilmark{13},
V.~M. Kaspi\altaffilmark{9}, 
M.~Koop\altaffilmark{7},
M.~T.~Lam\altaffilmark{2},
T.~J.~W.~Lazio\altaffilmark{8},
L.~Levin\altaffilmark{11},
A.~N.~Lommen\altaffilmark{5}, 
D.~R.~Lorimer\altaffilmark{11},
J.~Luo\altaffilmark{12},
R.~S.~Lynch\altaffilmark{14},
D.~R.~Madison\altaffilmark{2},
M.~A.~McLaughlin\altaffilmark{11}, 
S.~T.~McWilliams\altaffilmark{11},
C.~M.~F.~Mingarelli\altaffilmark{15, 16, 29},
D.~J.~Nice\altaffilmark{17},
N.~Palliyaguru\altaffilmark{11},
T.~T.~Pennucci\altaffilmark{18},
S.~M.~Ransom\altaffilmark{19}, 
L.~Sampson\altaffilmark{6},
S.~A.~Sanidas\altaffilmark{20, 21},
A.~Sesana\altaffilmark{22},
X.~Siemens\altaffilmark{4}, 
J.~Simon\altaffilmark{4},
I.~H.~Stairs\altaffilmark{10}, 
D.~R.~Stinebring\altaffilmark{23},
K.~Stovall\altaffilmark{24},
J.~Swiggum\altaffilmark{11},
S.~R.~Taylor\altaffilmark{8},
M.~Vallisneri\altaffilmark{8},
R.~van~Haasteren\altaffilmark{8,28},
Y.~Wang\altaffilmark{26}, \&
W.~W.~Zhu\altaffilmark{10, 16}}

\affil{$\star$Author order alphabetical by surname}
\affil{$^{1}$Center for Research and Exploration in Space Science and Technology and X-Ray Astrophysics Laboratory, \\NASA Goddard Space Flight 
Center, Code 662, Greenbelt, MD 20771, USA}
\affil{$^{2}$Department of Astronomy, Cornell University, Ithaca, NY 14853, USA}
\affil{$^{3}$National Radio Astronomy Observatory, 1003 Lopezville Rd., Socorro, NM 87801, USA}
\affil{$^{4}$Center for Gravitation, Cosmology and Astrophysics, Department of Physics, University of Wisconsin-Milwaukee,\\ P.O. Box 413, Milwaukee, WI 53201, USA}
\affil{$^{5}$Department of Physics and Astronomy, Franklin \& Marshall College, P.O. Box 3003, Lancaster, PA 17604, USA}
\affil{$^{6}$Department of Physics, Montana State University, Bozeman, MT 59717, USA}
\affil{$^{7}$Department of Astronomy and Astrophysics, Pennsylvania State University, University Park, PA 16802, USA}
\affil{$^{8}$Jet Propulsion Laboratory, California Institute of Technology, 4800 Oak Grove Drive, Pasadena, CA 91109, USA}
\affil{$^{9}$Department of Physics, McGill University, 3600  University St., Montreal, QC H3A 2T8, Canada}
\affil{$^{10}$Department of Physics and Astronomy, University of British Columbia, 6224 Agricultural Road, Vancouver, BC V6T 1Z1, Canada}
\affil{$^{11}$Department of Physics and Astronomy, West Virginia University, P.O. Box 6315, Morgantown, WV 26506, USA}
\affil{$^{12}$Center for Gravitational Wave Astronomy, University of Texas at Brownsville, Brownsville, TX 78520, USA}
\affil{$^{13}$Department of Physics, Columbia University, New York, NY 10027, USA}
\affil{$^{14}$National Radio Astronomy Observatory, P.O. Box 2, Green Bank, WV 24944, USA}
\affil{$^{15}$TAPIR, MC 350-17, California Institute of Technology, Pasadena, CA 91125, USA}
\affil{$^{16}$Max Planck Institute for Radio Astronomy, Auf dem H\"{u}gel 69, D-53121 Bonn, Germany}
\affil{$^{17}$Department of Physics, Lafayette College, Easton, PA 18042, USA}
\affil{$^{18}$University of Virginia, Department of Astronomy, P.O. Box 400325, Charlottesville, VA 22904, USA}
\affil{$^{19}$National Radio Astronomy Observatory, 520 Edgemont Road, Charlottesville, VA 22903, USA}
\affil{$^{20}$Anton Pannekoek Institute for Astronomy, University of Amsterdam, Science Park 904, 1098 XH ,Amsterdam, The Netherlands}
\affil{$^{21}$Jodrell Bank Centre for Astrophysics, University of Manchester, Manchester, M13 9PL, United Kingdom}
\affil{$^{22}$School of Physics \& Astronomy, University of Birmingham, Birmingham, B15 2TT, UK}
\affil{$^{23}$Department of Physics and Astronomy, Oberlin College, Oberlin, OH 44074, USA}
\affil{$^{24}$Department of Physics and Astronomy, University of New Mexico, Albuquerque, NM 87131, USA}
\affil{$^{25}$Department of Physics, Hillsdale College, 33 E. College Street, Hillsdale, Michigan 49242, USA}
\affil{$^{26}$School of Physics, Huazhong University of Science and Technology, 1037 Luoyu Road, Wuhan, Hubei Province 430074, China}

\affil{$^{27}$ Jansky Fellow}
\affil{$^{28}$ Einstein Fellow}
\affil{$^{29}$ Marie Curie Fellow}

\begin{abstract}
We compute upper limits on the nanohertz-frequency isotropic stochastic gravitational wave background (GWB) using the 9-year data release from the North American Nanohertz Observatory for Gravitational Waves (NANOGrav) collaboration. Well-tested Bayesian techniques are used to set upper limits for a GWB from supermassive black hole binaries under power law, broken power law, and free spectral coefficient GW spectrum models. We place a 95\% upper limit on the strain amplitude (at a frequency of yr$^{-1}$) in the power law model of $A_{\rm gw} < 1.5\times 10^{-15}$. We find that the upper limit on the individual spectral components is consistent with a white noise spectrum at frequencies greater than the inverse observation time. For a broken power law model, we place priors on the strain amplitude derived from simulations of \cite{s13} and \cite{mop14}. Using Bayesian model selection we find that the data favor a broken power law to a pure power law with odds ratios of 22 and 2.2 to one for the McWilliams and Sesana prior models, respectively. The McWilliams model is essentially ruled out by the data, and the Sesana model is in tension with the data under the assumption of a pure power law. Using the broken power-law analysis we construct posterior distributions on environmental factors that drive the binary to the GW-driven regime including the stellar mass density for stellar-scattering, mass accretion rate for circumbinary disk interaction, and orbital eccentricity for eccentric binaries, marking the first time that the \emph{shape} of the GWB spectrum has been used to make astrophysical inferences. We then place the most stringent limits so far on the energy density of relic GWs, $\Omega_\mathrm{gw}(f)\,h^2 < 4.2 \times 10^{-10}$, yielding a limit on the Hubble parameter during inflation of $H_*=1.6\times10^{-2}~m_{Pl}$, where $m_{Pl}$ is the Planck mass. Our limit on the cosmic string GWB, $\Omega_\mathrm{gw}(f)\, h^2 < 2.2 \times 10^{-10}$, translates to a conservative limit of $G\mu<3.3\times 10^{-8}$-- a factor of 4 better than the joint Planck and high-$l$~Cosmic Microwave Background data from other experiments.

\end{abstract}
\keywords{
Gravitational waves --
Methods:~data analysis --
Pulsars:~general --
}

\section{Introduction}
\label{sec:intro}


%
%
%
%
%
%

The 1967 discovery of pulsars \citep{hbp+68} inaugurated the age of high-energy astrophysics, and in time it led to the indirect but compelling confirmation that compact binaries lose energy to gravitational waves (GWs) in accordance with general-relativistic predictions \citep{tw89}.
%
%
%
It is therefore quite fitting that pulsars should now feature prominently in the quest to directly detect low-frequency GWs, and to use them as probes of the unseen, dynamical Universe. In this article, we describe how one can make astrophysical inferences about the underlying GW source population using the shape of the nanoHertz GW spectrum. Using the 9-year NANOGrav data release,~\citet{abb+15b}, we provide the first constraints on environmental factors which may be contributing the measured shape of the spectrum. We expect these novel techniques to become standard analysis tools, and as such, provide detailed descriptions of our methods. We also provide a summary of our results here, so that one may get an overview of our new methods and results before delving into the details.
\subsection{Gravitational Wave Detection with Pulsar Timing Arrays}
\citet{saz78} and \citet{det79} realized that GWs could manifest as otherwise unexplained residuals in the times of arrival (TOAs) of pulsar signals after subtracting a deterministic \emph{timing model}. This timing model accounts for the intrinsic evolution of pulsar spin, radio frequency-dependent delays to to interstellar propagation effects, the astrometric time delays and advances due to the relative motion of the pulsar system with respect to the Earth (indeed, to the observatory), as well as orbital-kinematic and light-propagation effects for pulsars that orbit a binary companion (see, e.g., \citealt{handbook}). \citet{fb90} pointed out that the timing residuals from an array of pulsars (a \emph{pulsar timing array}, or PTA) could be analyzed coherently to separate GW-induced residuals, which have distinctive correlations among different pulsars \citep{hd83}, from other systematic effects, such as clock errors or delays due to light propagation through the interstellar medium.

Today, three international consortia [NANOGrav \citep{ml13}, the EPTA \citep{kc13}, and the PPTA \citep{h13}] are more than ten years into extensive campaigns to search for GWs by timing dozen of individual millisecond pulsars (MSPs), in which the best-timed have rms residuals less than 100 ns (corresponding to GW strain sensitivities $\sim 10^{-15}$). The three PTAs collaborate and share data under the aegis of the International Pulsar Timing Array \citep{haa+10}.

In order to robustly detect GWs, one must have a thorough understanding of the underlying noise in the pulsar timing data \citep[see e.g.][for a detailed review]{c13, s13d}.
Template matching errors due to radiometer noise are uncorrelated in both time and frequency, but pulse-jitter noise
\citep{cs10} appears to affect all TOAs obtained simultaneously in different frequency channels.  Correlated timing noise with a red power spectrum occurs to varying degree in different pulsars.  Spin noise \citep{sc10} is achromatic and is much smaller in MSPs compared to objects with stronger magnetic fields and longer spin periods.  Chromatic red noise due to propagation through intervening plasmas (ISM, interplanetary medium, and ionosphere) may also be present if dispersive delays are not removed perfectly or if scattering and refraction effects contribute significantly.  

\subsection{The stochastic GW background from SMBHBs}

PTAs are most sensitive to GWs with frequencies on the order of the inverse timespan of timing observations, where TOA measurement noise averages out most efficiently. The strongest expected sources in this band are supermassive black hole binaries (SMBHBs) with masses of $10^{8}\mbox{--}10^{10} \, M_\odot$, out to $z \simeq 1$ \citep{rm95,jb03,wl03}. The binaries form after the hierarchical mergers \citep{shm+04,svc08} of galaxies hosting individual SMBHs (as most galaxies are thought to do, cf.\ \citealt{kh13}). The most massive and nearest binaries may be detected individually by PTAs through their continuous GW emission. Moreover, the cosmic population of SMBHBs may be observed collectively as a stochastic GW background composed of the incoherent superposition of signals from the binaries. \citet{rsg15} discuss which detection is likely to come first.

The main focus of this article is the measurement of an isotropic stochastic GW background from SMBHBs. Anisotropic-background searches based on the formalism and techniques developed by \citet{grt+14,tg13,msmv13} are currently underway and will be the subject of a follow-up paper. We briefly consider stochastic backgrounds from relic (or primordial) GWs as well as backgrounds from cosmic (super)strings as complementary studies.
Our main focus is to demonstrate how, even in the absence of a positive detection, PTA data can be used to constrain and characterize the astrophysical processes and SMBHB source populations that give rise to the GW background.

The simplest characterization of the stochastic GW background -- a power-law Gaussian process with isotropic inter-pulsar correlations -- applies if:
\begin{enumerate}
\item all binaries are assumed to have circular orbits (so each component signal is instantaneously monochromatic);
\item all binaries evolve through the PTA band due purely to GW emission, as opposed to environmental effects such as interactions with nearby gas or with stars in the galactic nucleus;
\item all binaries are distributed isotropically across the sky in sufficient numbers to fulfill the central limit theorem at all frequencies.
\end{enumerate}
Under these conditions, the observed timing residuals due to the GW background are described fully by the (cross-) power spectral density
\begin{equation}
S_{ab}(f) = \Gamma_{ab} \times \frac{A_{\rm gw}^2}{12 \pi^2} \left(\frac{f}{\mathrm{yr}^{-1}} \right)^{-\gamma} \, \mathrm{yr}^3,
\label{eq:powerlaw}
\end{equation}
where $a$ and $b$ range over the pulsars in the array, $\gamma = 13/3$ for a background composed of SMBHBs, and $\Gamma_{ab}$ is the Hellings--Downs \citeyearpar{hd83} isotropic correlation coefficient for pulsars $a$ and $b$ (a function of the separation angle between their lines of sight, which is normalized so that $\Gamma_{aa} = 1$; see also \citealt{ms14}). 
Power-law GW backgrounds are also described (independently of observations) in terms of their \emph{characteristic strain}
\begin{equation}
h_c(f) = A_{\rm gw}\biggl(\frac{f}{\mathrm{yr}^{-1}}\biggr)^\alpha,
\label{eq:gwstrain}
\end{equation}
which is related to Eq.\ \eqref{eq:powerlaw} by $S_{ij}(f) = \Gamma_{ij} h_c(f)^2 / (12 \pi^2 f^3)$ and $\gamma = 3 - 2 \alpha$ ($\alpha=-2/3$ for SMBHBs). 

Recent predictions for the value of $A_\mathrm{gw}$, based on models of SMBH--galaxy coevolution and on observational constraints of galaxy assembly and SMBH mass functions, range between $\sim 10^{-15}$ and $10^{-14}$ (\citealt{mop14,s13, rws+14}---hereafter \citetalias{mop14}, \citetalias{s13}, and \citetalias{rws+14}). 
The \citetalias{mop14} model assumes that all SMBHBs are in circular orbits evolving under GW emission alone, as well as a single black hole--host correlation from \cite{mm13}, yielding an estimate of the stochastic GW background that is roughly four times as large as \citetalias{s13} and \citetalias{rws+14}. The \citetalias{s13} model uses a wide variety of galaxy merger rates and empirical black hole--host relations to yield a collection of phenomenological SMBHB merger rates, which are used to compute a distribution of possible GW signals. Note that for this paper we consider distributions from \citetalias{s13} that \emph{only} use the black hole-host galaxy relations of \cite{mm13} and \cite{gs13} The \citetalias{rws+14} model also assumes the black-hole-host host correlation of \cite{mm13} but includes the possibility of the SMBHBs evolving in stellar environments, and accounts for non-zero binary eccentricities. Thus, some of these models predict spectral densities that deviate from straight power-law behavior at low frequencies; in that case, we refer the fiducial $A_\mathrm{gw}$ to their value at a frequency of yr$^{-1}$. Finally, recent results, \citep{kh13}, indicate that the black hole-host correlation's normalization is being revised to larger values with more observations indicating that an even stronger GWB may be expected; however, for this work we use the published results based on \cite{mm13} to make the most fair comparison among models.

\subsection{New results in this paper}

Over the last few years, the three PTAs have reported ever improving upper limits on the GW backgrounds of the form \eqref{eq:powerlaw}: $A_\mathrm{gw} < 7 \times 10^{-15}$ (NANOGrav, \citealt{dfg+13}, hereafter \citetalias{dfg+13}), $6 \times 10^{-15}$ (EPTA, \citealt{vhj+11}), $3 \times 10^{-15}$ (EPTA, \citealt{ltm+15}, hereafter \citetalias{ltm+15}), $2.4 \times 10^{-15}$ (PPTA, \citealt{src+13}), all quoted at 95\% confidence and a reference frequency of yr$^{-1}$, although differences in the statistical analyses and in the availability and selection of pulsar datasets mean that these numbers are not entirely homogeneous. It is clear that there is significant tension between these observational limits and astrophysical expectations for $A_\mathrm{gw}$. It is important to note that a limit on $A_{\rm gw}$ does note translate \emph{directly} to a limit on the SMBHB population because of the finite number of pulsars that contribute to the limit and the stochasticity of the GW signal itself.

This statement can be made more precise. In Sec.\ \ref{sec:bayesian-analysis} of this paper we report a new 95\% upper limit $A_\mathrm{gw} < 1.5 \times 10^{-15}$, obtained from the Bayesian analysis of NANOGrav's 9-year, 37-pulsar dataset released in 2015 \citep{abb+15b}. This limit is five times more constraining than the same analysis applied to NANOGrav's 5-year dataset \citepalias{dfg+13} (see the top of Fig.\ \ref{fig:9-5-amplitude} for a comparison of the two posterior probability distributions). Now, following \citet{src+13}, we can assess the consistency of our result with astrophysical GW-background models. We find a 0.8\% probability that the observed $A_\mathrm{gw}$, as characterized probabilistically by its Bayesian posterior, is drawn from the amplitude distribution developed in \citetalias{mop14}, and a 20\% probability that it is drawn from the (very similar) \citetalias{rws+14} and \citetalias{s13} distributions. Correspondingly, the two bottom panels of Fig.\ \ref{fig:9-5-amplitude} show that 9-year observations update significantly the \citetalias{mop14} and \citetalias{rws+14}/\citetalias{s13} amplitude priors, much more so than our 5-year dataset.

In Sec.\ \ref{sec:frequentist-analysis} we report also a frequentist, \emph{optimal-statistic} 95\% upper limit $A_\mathrm{gw} < 1.3 \times 10^{-15}$, a fivefold improvement on the analogous result of \citetalias{dfg+13}; however, the optimal statistic is problematic in the presence of marginally-detectable GW signals,  so we offer it only as a proxy for the improving sensitivity of NANOGrav's observations.

Stochastic GW backgrounds in the PTA band may also originate from quantum fluctuations amplified during inflation \citep{g05} and from topological broken-symmetry remnants such as  \citep{dv01,oms13}, for which Eq.\ \eqref{eq:powerlaw} applies with $\gamma = 5$ (depending on the equation of state $w$) and $16/3$, respectively. 

In Sec.\ \ref{sec:relics} we obtain 95\% upper limits $A_{\rm gw} < 8.1\times 10^{-16}$ for relic GWs, corresponding to energy-density limits $\Omega_\mathrm{gw}(f=1/\mathrm{yr})\,h^2 < 4.2 \times 10^{-10}$, where $h$ parametrizes the Hubble constant $H_0 \equiv h \times 100 \, \mathrm{km}/\mathrm{s}/\mathrm{Mpc}$. This limit is a factor of 2.9 better than limit reported by the EPTA in \citet{ltm+15}. We then obtain limits on the Hubble parameter during inflation, $H_*=1.6\times10^{-2}~m_{Pl}$, where $m_{Pl}\equiv 1/\sqrt{G}$ is the Planck mass, using the method developed by \cite{z11}. 

In Sec.\ \ref{sec:strings} cosmic strings, we find $A_{\rm gw} < 6\times 10^{-16}$ and $\Omega_\mathrm{gw}(f=1/\mathrm{yr})\, h^2 < 2.23 \times 10^{-10}$, corresponding to a conservative limit on the string tension of $G\mu<3.3\times 10^{-8}$ -- a factor of 4 better than the joint Planck and high-$l$~Cosmic Microwave Background data from other experiments. If we then restrict ourselves to a GWB produced by the production of large cosmic string loops, as described by \cite{bjo+14}, then our string tension limit is much more restrictive: $G\mu<1.3\times10^{-10}$, a factor of 6.6 times more constraining than an identical analysis performed using the EPTA limit.

\subsection{Astrophysical Inference}

The mismatch between observations and expectations can be explained in different ways.
First, it is possible that the astrophysical models and the three assumptions listed above \emph{are} correct, but the background amplitude realized in nature lies in the tails of the predicted distributions. This hypothesis obviously wanes as upper limits get more stringent.

Second, it is possible that some of the \emph{inputs} of the astrophysical models are not estimated correctly; we can then use GW observations to constrain these inputs. For example, in Sec.\ \ref{sec:joesarah} we assume measurements of the galaxy mass function and merger rate, and we constrain the scaling between the galaxy bulge mass and the central SMBH mass, which affects the observed $A_\mathrm{gw}$ most significantly, through the distribution of binary chirp masses. We find a slight inconsistency between the scaling relation reported by \citet{kh13} and our limit, whereas the \cite{mm13} estimate is consistent within its margin of error. 
The \citet{mm13} black hole--stellar velocity dispersion relation underlies the \citetalias{mop14} predictions, while \citetalias{s13} and \citetalias{rws+14} take into account a variety of alternative black hole--host estimates. 

Third, the simple GW-background characterization that yields Eq.\ \eqref{eq:powerlaw} may not be realistic, because SMBHBs may form with significant eccentricity and retain it into the PTA band, distributing GW emission over a range of frequency harmonics. Furthermore, environmental effects (interactions with stars on centrophilic orbits in galactic nuclei, or with circumbinary gas disks) can accelerate the transit of individual binary systems through the PTA band (see Secs.\ \ref{sec:eccentricity} and \ref{sec:environs} for details and references). These environmental effects deplete the GW background at low frequencies where PTA measurements are most sensitive (i.e., frequencies $\sim$ the inverse observation timespan), so PTA upper limits may yet be compatible with the \citetalias{mop14}/\citetalias{s13}/\citetalias{rws+14} predictions at the higher frequencies where the $\gamma = 13/3$ power law is realized.

To investigate this point, in Sec.\ \ref{sec:AstrophysicalModel} we reanalyze the NANOGrav data using a phenomenological $S_{ij}(f)$ in the form of an inflected power law, parametrized by a turnover frequency $f_\mathrm{bend}$ and by a shape parameter $\kappa$, as proposed by \citet{scm15} [see Eq.\ \eqref{eq:turnover}]. By combining this enhanced GW-background model with \citetalias{mop14} and \citetalias{s13}/\citetalias{rws+14} amplitude priors, we conclude that the data prefer an inflected spectrum to a moderate degree for \citetalias{mop14}, however this preference is very weak for \citetalias{s13}/\citetalias{rws+14} models. Quantitatively, the Bayes factors between enhanced and pure-power-law spectral models for each of the two priors are $22$ and 2.2, respectively; graphically, the shading in Fig.\ \ref{fig:bayesogram} represents the frequency-by-frequency posterior probability density for the GW spectrum, which appears significantly inflected for \citetalias{mop14}, and only slightly so for \citetalias{s13}/\citetalias{rws+14}.
The data are not sufficiently informative to constrain the amplitude and shape of the spectrum jointly in the absence of a compact prior, so we cannot produce a unique metric of consistency, as we did above in the case of the simple power-law spectrum.

Beyond this phenomenological characterization, the joint $(A_\mathrm{gw},f_\mathrm{bend},\kappa)$ posteriors can be mapped into constraints for the SMBHB eccentricities (which we do in Sec.\ \ref{sec:eccentricity}) and for the astrophysical variables that govern environmental interactions (Sec.\ \ref{sec:environs}). For the former, we assume for simplicity that all binaries had the same eccentricity $e_0$ when the semi-major axis of their orbits was 0.01 pc and that they evolved purely by GW emission since; we follow \citet{hmgt15} to construct eccentric binary populations and GW strain spectra. The resulting posteriors on $e_{0}$ indicate that  $e_0 \gtrsim 0.7$ is preferred for the \citetalias{mop14} prior, and $e_0 \gtrsim 0.5$ is preferred for \citetalias{s13}/\citetalias{rws+14} (though still consistent with smaller values). These limits suggest that either SMBHBs form with rather high $e_0$, or that binary eccentricity is not a good explanation for the mismatch between $A_\mathrm{gw}$ observations and predictions.

To characterize environmental interactions (see Sec.\ \ref{sec:environs}), we compute the evolution of orbital frequency due to stellar scattering events and to circumbinary gas disk interactions as $\mathrm{d}f/\mathrm{d}t \propto f^{1/3}$ and $f^{4/3}$, respectively, corresponding to $\kappa = 10/3$ and $7/3$ (since for GW-driven evolution $\mathrm{d}f/\mathrm{d}t \propto f^{11/3}$); the frequency $f_\mathrm{bend}$ then marks the transition between environmentally- and GW-driven evolution. For the case of stellar scattering, $f_\mathrm{bend}$ depends most significantly on the mass density $\rho$ of galactic-core stars. Astrophysical estimates for $\rho$ are quite uncertain
with typical values between $10-10^4~\Msun$~pc$^{-3}$~\cite{dch+07} assuming that a majority of our GW sources come from merging elliptical galaxies. Under several simplifying assumptions (e.g., that all binaries have circular orbits, that all galaxies have comparable densities, and that only a single environmental effect is active), we find that $\rho \gtrsim 10^4 \, M_\odot \, \mathrm{pc}^{-3}$ is strongly preferred for the \citetalias{mop14} amplitude prior, and the data is unconstraining for the \citetalias{s13}/\citetalias{rws+14} prior. 
For the circumbinary disk case, $f_\mathrm{bend}$ depends on the accretion rate on to the primary (most massive) BH, $\dot M_1$. The accretion rate of the primary BH is a function of $M_1$, the mass of the primary BH, $\epsilon$, the radiative efficiency parameter with a canonical value of $\epsilon=0.1$ and $\kappa_\mathrm{opp}$ is the disk opacity, $\dot M_1 \propto M_1 \epsilon^{-1} \kappa^{-1}_\mathrm{opp}$. Hence $\dot M_1$ takes on a range of values, typically $10^{-3}\Msun$~yr$^{-1}$ -- $1~\Msun$~yr$^{-1}$, see e.g. \citealt{mcf01, an02, gcs+15}.
In our analysis, we find that the accretion rate  $\dot{M}_1 \gtrsim 10^{-1} \, M_\odot \, \mathrm{yr}^{-1}$ is strongly preferred for the \citetalias{mop14} amplitude prior, and again, the data is unconstraining for the \citetalias{s13}/\citetalias{rws+14} prior.

\subsection{Plan of the Paper}
This paper presents the first analysis that characterizes the spectral amplitude \emph{and shape} of the GW background; in some cases we find tension between observations and predictions for the GW background from SMBHBs. When informed with astrophysical amplitude priors, the data favor phenomenological models that include an inflection point over pure GW-driven power laws.  We attempt to interpret the phenomenological posteriors in terms of binary-eccentricity or environmental effects by carrying out our analyses in sequence, investigating different effects \emph{separately}. Admittedly, we rely on  simple analytical models of eccentricity and environments; to some extent, this simplicity becomes necessary if we are to draw any astrophysical conclusion from data that are not yet very informative. Therefore, it will be extremely interesting to improve the sophistication of our analysis, and to apply it to longer, richer datasets, such as the upcoming NANOGrav 11-year dataset, as well as the multiple-PTA datasets assembled by the International Pulsar Timing Array.

The rest of this paper is organized as follows: in Sec.\ \ref{sec:obs} we describe the NANOGrav 9-year pulsar-timing and dataset; in Sec.\ \ref{sec:data_analysis} we discuss our signal and noise models, as well as the statistical framework of our analysis; in Sec.\ \ref{sec:results} we document important implementation details, and report our results in detail; in Sec.\ \ref{sec:discussion} we discuss the astrophysical interpretation of our analysis, and derive limits on astrophysical and cosmological quantities; in Sec.\ \ref{sec:conclusions} we offer our conclusions.

\section{Observations}
\label{sec:obs}

This paper uses the observations from the NANOGrav 9-year data release, recently
presented in \citet{abb+15b}, which contains observations made over a time span
from 2004 to 2013. Initially the array consisted of 15 pulsars, and it grew to
37 pulsars over the course of the project. The first five years of data on 17
pulsars constituted the NANOGrav 5-year data release, previously reported by
\citetalias{dfg+13}. In this release, all data have been reprocessed. We give a brief
overview of the dataset in this section; for details we refer the reader to
\citet{abb+15b}.

\subsection{Observatories} \label{sec:observatories}
The 305\,m William E.\ Gordon Telescope of the Arecibo Observatory (AO) was used
to observe pulsars with declinations in the range $0^\circ<\delta<39^\circ$;
pulsars outside of this range were observed with the 100\,m Robert C.\ Byrd
Green Bank Telescope (GBT) of the National Radio Astronomy Observatory (NRAO). 
The pulsars PSRs J1713+0747 and B1937+21 were observed at both AO and the GBT. 
The typical observation cadence was about once every month.

 At AO, four receivers were used: 327 MHZ, 430 MHz, 1400 MHz, and 2100 MHz. Of
those, typically two (or more) were used in immediate succession within $\sim$1
hour per observation session, which is possible at AO since the receiver does
not need to be physically replaced for a receiver change. At the GBT, 
observations were made with two receivers at 800 MHz and 1400 MHz.
Observations were only included in the dataset if observations could be made
within a time span of 14~days; otherwise the observations were discarded for the
lack of information about variations in the interstellar medium dispersion.

Two sets of nearly-identical data acquisition systems were developed
specifically for NANOGrav. Early observations through 2012.3 (AO) and 2011.0
(GBT) were recorded by the Astronomical Signal Processor (ASP) and the Green
Bank Astronomical Signal Processor (GASP) respectively \citep{dem07}. The later
observations beginning at 2012.2 (AO) and 2010.2 (GBT) were recorded using the
Puerto Rican Ultimate Pulsar Processing Instrument (PUPPI) and the Green Bank
Ultimate Pulsar Processing Instrument (GUPPI) respectively
\citep{drd+08,fdr10}. These instruments performed real-time coherent
dedispersion on the digitized incoming baseband data, and folded the data using
a pre-computed ephemeris. After RFI excision, polarization calibration, and flux
calibration, the end product of each instrument consisted of total-intensity
pulse profiles for a series of frequency channels. These profiles were
integrated over the course of an observation, resulting in one or more
subintervals of typically 20-30 minutes each.

\subsection{Time of Arrival data} \label{sec:toas}
TOAs were created using various tools in the \psrchive package: the
Fourier-domain algorithm of \cite{tay92} to calculate TOAs, and denoising the pulse
profiles via wavelet decomposition. Offsets resulting from latencies between
different observing systems were determined as overall
timing-model-fit-parameters, which were taken into account when doing the timing
noise analysis.

For each pulsar, TOAs were calculated for all frequency channels recorded from a
given receiver. The effect of time-varying dispersion was taken into account by
including ``DMX'' parameters in the timing model \citepalias{dfg+13}, which essentially allows for
an extra delay proportional to $1/\nu^2$ to be fit for, where $\nu$ is radio
frequency.
Pulse shape evolution with frequency is taken into account differently than in
\citetalias{dfg+13}. Instead of including a phase offset per frequency channel, a
heuristic mitigation procedure is used that parameterizes the profile evolution
with frequency, the details of which are in \citet{abb+15b}.

In summary, similar to the NANOGrav 5-year release, the NANOGrav 9-year release
consists of high-quality, publicly available\footnote{\url{http://data.nanograv.org}} TOAs for 37~pulsars, produced per
frequency sub-channel.

\section{Data Analysis Methods}
\label{sec:data_analysis}

All the data analysis methods we use in this manuscript (both the Bayesian and frequentist methods) are effectively carried out in the time-domain. We start in Sec.~\ref{sec:likelihood} by defining the likelihood function and introducing our notation, which follows that of \citet{abb+15b}.
We continue our discussion with the Bayesian approach in Sec.~\ref{sec:bayes}, and the frequentist approach in the form of the optimal cross-correlation statistic in Sec.~\ref{sec:optimalstatistic}.

\subsection{Likelihood} \label{sec:likelihood}
We start our discussion
by decomposing our $N_{\rm TOA}$ timing residual data for a single pulsar $\delta\mathbf{t}$ in its
individual constituents as follows:
\begin{equation}
\delta\mathbf{t} = M\boldsymbol{\epsilon} + F\mathbf{a} + U\mathbf{j} + \mathbf{n}.
\label{eq:timingresiduals}
\end{equation}
The term $M\boldsymbol{\epsilon}$ describes inaccuracies in the subtraction of
the timing model, where $M$ is called the timing model design matrix, and
$\boldsymbol{\epsilon}$ is a vector of small timing model parameter offsets. The
term $F\mathbf{a}$ describes all low-frequency signals, including low-frequency
(``red'') noise,
with a limited number of Fourier coefficients $\mathbf{a}$. Our harmonics are chosen as integer multiples of the harmonic base frequency
$1/T$, with $T$ the length of our dataset (either of a single pulsar, or the
entire array of pulsars). The matrix $F$ then has alternating sine and cosine
functions. We note that this is just a particular choice of rank-reduced basis,
and we could have chosen many others without influencing our results. The term $U\mathbf{j}$ describes noise that is
fully correlated (correlation coefficient of $1$) for simultaneous observations at different observing
frequencies, but fully uncorrelated in time. The matrix $U$ is an $N_{\rm
TOA}\times N_{\rm epoch}$ matrix that maps the $N_{\rm epoch}$ observation sessions to the $N_{\rm
TOA}$ TOAs.
The vector $\mathbf{j}$ describes the white noise per
observation session that is fully correlated across all observing frequencies.
The last term, $\mathbf{n}$, describes Gaussian white noise that is assumed to be
left in the data, after correcting for all known systematics.

The white noise is assumed to be an uncorrelated, Gaussian noise process, with
variance-covariance:
\begin{equation}
    \langle \mathbf{n}_{i}\mathbf{n}_{j}\rangle = \sum_{k} N_{ij,k} = \sum_{k} E_{k}^2(W_{ij}+Q_{k}^2\delta_{ij}),
\end{equation}
where $E_k$ and $Q_k$ are the {\sc TEMPO} and {\sc TEMPO2} ``EFAC'' and ``EQUAD'' parameters for each observing backend $k$, $\delta_{ij}$ is the Kronecker delta function, and $W={\rm diag}\{\sigma_i^2\}$ are the TOA uncertainties. The matrix elements $(i,j)$ apply to the TOAs
corresponding to the observing system labeled by $k$. In practice we cannot fully separate various contributions to our TOAs, so we have to take into account that our corrections for the various terms of Eq.~\eqref{eq:timingresiduals} are inaccurate. We do this by constructing our likelihood from the noise-mitigated timing residuals:
\begin{equation}
    \mathbf{r} = \delta\mathbf{t}-M\mathbf{\epsilon}-F\mathbf{a}-U\mathbf{j},
\end{equation}
where $\mathbf{r}$ is our best approximation of $\mathbf{n}$, given our knowledge of all the noise and signal parameters. The likelihood can now be written down as:
\begin{equation}
    p(\delta\mathbf{t}|\boldsymbol{\epsilon}, \mathbf{a}, \mathbf{j},\mathbf{\phi}) = \frac{\exp\left( -\frac{1}{2}\mathbf{r}^TN^{-1}\mathbf{r} \right)}{\sqrt{\det(2\pi N)}},
    \label{eq:expandedlikelihood}
\end{equation}
where $N=\sum_{k}N_{k}$ represents the total effect of all white noise. We have collectively denoted all parameters not directly represented by $\boldsymbol{\epsilon}$, $\mathbf{a}$, and $\mathbf{j}$ as $\mathbf{\phi}$. Henceforth, we shall refer to $\boldsymbol{\epsilon}$ as the hyperparameters.
We group the reduced-rank signals as follows:
\begin{equation}
    T = \bb M  & F  & U \eb, \quad
    \mathbf{b} = \bb \boldsymbol{\epsilon} \\ \mathbf{a} \\ \mathbf{j} \eb,
\end{equation}
which allows us to elegantly place a Gaussian prior on the coefficients of these
random processes. The prior covariance is:
\begin{equation}
    B = \bb \infty & 0 & 0 \\ 0 & \varphi & 0 \\ 0 & 0 & \msJ \eb,
\end{equation}
resulting in a prior:
\begin{equation}
    p(\mathbf{b}|\boldsymbol{\phi}) = \frac{\exp\left(
    -\frac{1}{2}\mathbf{b}^{T}B^{-1}\mathbf{b} \right)}{\sqrt{\det(2\pi B)}},
    \label{eq:expandedprior}
\end{equation}
where $\infty$ is a diagonal matrix of infinities, which effectively means we
have a uniform unconstrained prior on the timing model parameters
$\mathbf{\epsilon}$. As described in \citet{abb+15b}, this representation allows
us to analytically marginalize Eq. \eqref{eq:expandedlikelihood} times Eq. \eqref{eq:expandedprior}
over the waveform coefficients $\mathbf{b}$ of
the noise, resulting in a drastic reduction of the dimensionality of our
posterior \citep{lha+13,vhv14,vhv15}:
\begin{equation}
    p(\delta\mathbf{t}|\boldsymbol{\phi}) = \frac{\exp\left(
    -\frac{1}{2}\delta\mathbf{t}^TC^{-1}\delta\mathbf{t}
    \right)}{\sqrt{\det(2\pi C)}},
    \label{eq:reducedlikelihood}
\end{equation}
with $C = N + TBT^{T}$. The Woodbury matrix identity \citep{w50} can be used to
evaluate Eq.~\eqref{eq:reducedlikelihood} efficiently.

The parameters that describe $B$ are the hyperparameters $\boldsymbol{\phi}$.
The
hyperparameters of the diagonal matrix $\msJ$ are the per-backend {\sc TEMPO2} ECORR
parameters. The matrix $\varphi$ represents the spectrum of the low-frequency noise
and the stochastic gravitational waves, and it therefore contains terms
correlated between pulsars. Denoting pulsar number with
$(a,b)$, and frequency bin with $(i,j)$, we can write:
\begin{equation}
    [\varphi]_{(ai),(bj)}=\Gamma_{ab}\rho_i\delta_{ij} +
    \eta_{ai}\delta_{ab}\delta_{ij},
\end{equation}
where $\eta_{a}$ is the spectrum of the low-frequency noise of pulsar $a$, $\rho$ is the spectrum of the GWB, and $\Gamma_{ab}$ is the signal correlation matrix. The elements of the signal correlation matrix consist of the overlap reduction function for a GWB signal, which is a dimensionless function that quantifies the correlated response of the pulsars to a stochastic GWB \citep{ms14}.
The quantities $\rho_i$ and $\eta_{ai}$ can
be either modeled as independent model parameters (i.e., per frequency), or as
modeled spectral density with a specific shape (e.g., a power-law model).
We note that $\varphi$ can be represented by a block-diagonal matrix, where each
block corresponds to a specific frequency bin; all frequencies are theoretically
independent degrees of freedom.


\subsection{Bayesian analysis} \label{sec:bayes}

As an alternative to the orthodox frequentist approach to data analysis, Bayesian inference is a method of statistical inference in which Bayes rule of conditional probabilities is used to update one's knowledge as observations are acquired. Given a model $\mathcal{H}$, model parameters $\mathcal{\Theta}$, and observations $\mathcal{D}$, we write Bayes rule as:
\begin{equation}
    Pr\left(\mathcal{\Theta}|\mathcal{D},\mathcal{H}\right)
    Pr\left(\mathcal{D}|\mathcal{H}\right) =
    Pr\left(\mathcal{D}|\mathcal{\Theta},\mathcal{H}\right)
    Pr\left(\mathcal{\Theta}|\mathcal{H}\right),
    \label{eq:bayes}
\end{equation}
where $Pr\left(\mathcal{\Theta}|\mathcal{D},\mathcal{H}\right) \equiv Pr\left(\mathcal{\Theta}\right)$ is the posterior (probability distribution) of the parameters, $Pr\left(\mathcal{D}|\mathcal{\Theta},\mathcal{H}\right) \equiv L\left(\mathcal{\Theta}\right)$ is the likelihood, $Pr\left(\mathcal{\Theta}|\mathcal{H}\right) \equiv \pi\left(\mathcal{\Theta}\right)$ is the prior (probability distribution), and
$Pr\left(\mathcal{D}|\mathcal{H}\right) \equiv \mathcal{Z}$ is the marginal likelihood or evidence.

The left-hand side of Eq.~\eqref{eq:bayes} can be regarded as the ``output'' of the Bayesian analysis, and the right-hand side is the ``input''. Indeed, provided we have a generative model of our observations (meaning we can simulate data, given the model parameters), we know the likelihood and prior. However, for parameter estimation we would like to know the posterior, and for model selection we need the evidence.

For parameter estimation, the evidence $\mathcal{Z}$ is usually ignored, and one can use $L(\mathcal{\Theta})\pi(\mathcal{\Theta})$ directly to estimate confidence intervals. Typically one provides confidence intervals for single components and pairs of elements of $\mathcal{\Theta}$. This involves an integral over $Pr(\mathcal{\Theta})$ over all but one or two parameters, a process called marginalization.
When $\Theta$ is higher-dimensional, Monte-Carlo sampling methods are typically
used to perform this multi-dimensional integral. We use Markov Chain Monte Carlo methods in this work to sample the posterior distribution.

Model selection between two models $\mathcal{H}_0$ and $\mathcal{H}_1$ can be carried by calculating the ``Bayes factor'': the ratio between the evidence for the two models. Assuming we have a prior degree of belief of how likely the two model are ($Pr(\mathcal{H}_0)$ and $Pr(\mathcal{H}_1)$), we can write:
\begin{equation}
    O = \frac{Pr(\mathcal{H}_1|\mathcal{D})}{Pr(\mathcal{H}_0|\mathcal{D})} =
    \mathcal{B}_{10}(\mathcal{D}) \frac{Pr(\mathcal{H}_1)}{Pr(\mathcal{H}_0)},
    \label{eq:oddsratio}
\end{equation}
where $\mathcal{B}_{10}(\mathcal{D}) \equiv \mathcal{Z}_1/\mathcal{Z}_0$ is the Bayes factor, and $O$ is the odds ratio. The odds ratio can be obtained by calculating the evidence $\mathcal{Z}$ for each model separately (e.g. with Nested Sampling or thermodynamic integration), or by calculating the Bayes factor $\mathcal{B}$ between two models directly (e.g. with transdimensional markov chain Monte Carlo methods).


\subsection{Optimal cross-correlation statistic} \label{sec:optimalstatistic}

The \textit{optimal statistic} \citep{abc+09,dfg+13,ccs+15} is a frequentist estimator of the isotropic GWB strain-spectrum amplitude in the weak-signal regime. The estimator maximizes the likelihood of Eq.~\eqref{eq:reducedlikelihood}, and it can be written as:
\begin{equation}
\label{eq:os}
\hat{A}_{\rm gw}^2 = \frac{\sum_{ab}\delta\mathbf{t}_a^{\rm T}\mathbf{P}_a^{-1}\tilde{\mathbf{S}}_{ab}\mathbf{P}_b^{-1}\delta\mathbf{t}_b}{\sum_{ab}{\rm tr}\left[\mathbf{P}_a^{-1}\tilde{\mathbf{S}}_{ab}\mathbf{P}_b^{-1}\tilde{\mathbf{S}}_{ba}\right]},
\end{equation}
where $\mathbf{P}_a = \langle\delta\mathbf{t}_a\delta\mathbf{t}_a^{\rm T}\rangle$ is auto-covariance matrix of the residuals of a single pulsar. This is $C$ of Eq.~\eqref{eq:reducedlikelihood} for a noise-only model.
The term $A_{\rm gw}^{2}\tilde{\mathbf{S}}_{ab}=\mathbf{S}_{ab}=\langle \delta\mathbf{t}_a\delta\mathbf{t}_b^{\rm T} \rangle$ represents the signal covariance between pulsar $a$ and pulsar $b$. As described in the previous section, our model contains no other signals with a non-zero correlation coefficient between different pulsars. The normalization of the optimal statistic is such that $\langle\hat{A}_{\rm gw}^2\rangle = A^{2}_{\rm gw}$.

Following \citet{ccs+15}, we also quote expressions for the variance, and the signal-to-noise ratio (S/N, in power) of the optimal statistic. The variance of the optimal statistic in the absence of a GWB signal is given by:
\begin{equation}
    \sigma_0^2 = \left(\sum_{ab}{\rm tr}\left[\mathbf{P}_a^{-1}\tilde{\mathbf{S}}_{ab}\mathbf{P}_b^{-1}\tilde{\mathbf{S}}_{ba}\right]\right)^{-1}.
    \label{eq:osvariance}
\end{equation}
Although this expression is not valid in general, in the weak-signal regime, in which the cross correlated power is ignored from the variance, this can be used as an approximation of the variance in $\hat{A}_{\rm gw}^2$. The S/N for a given signal and noise realization is:
\begin{equation}
\rho = \frac{\hat{A}_{\rm gw}^2}{\sigma_0} = \frac{\sum_{ab}\delta\mathbf{t}_a^{\rm T}\mathbf{P}_a^{-1}\tilde{\mathbf{S}}_{ab}\mathbf{P}_b^{-1}\delta\mathbf{t}_b}{\left(\sum_{ab}{\rm tr}\left[\mathbf{P}_a^{-1}\tilde{\mathbf{S}}_{ab}\mathbf{P}_b^{-1}\tilde{\mathbf{S}}_{ba}\right]\right)^{1/2}},
\end{equation}
which has an expectation value of
\begin{equation}
\langle\rho\rangle = A_{\rm gw}^2\left(\sum_{ab}{\rm tr}\left[\mathbf{P}_a^{-1}\tilde{\mathbf{S}}_{ab}\mathbf{P}_b^{-1}\tilde{\mathbf{S}}_{ba}\right]\right)^{1/2}.
\end{equation}
These expressions generalize the detection significance estimator provided in \citet{jhl+05}, properly taking into account the spectrum of the signal and the noise, as well as details such as the irregular sampling. The S/N here has the same meaning, which if interpreted as a zero-mean unit-variance normal distribution, can be used to place upper-limits on the GWB amplitude. Clearly, this interpretation is only appropriate in the weak-signal regime, but it serves as an independent sanity check for our other methods. Setting our statistical significance threshold $\alpha=0.95$, we can place a one-sided upper-limit as:
\begin{equation}
    A^2_{\rm ul} = \hat{A}_{\rm gw}^2 + \sqrt{2}\sigma_0{\rm erfc}^{-1}[2(1-\alpha)].
    \label{eq:osupperlimit}
\end{equation}
We note that all the usual caveats of frequentist-type upper-limits apply to this methodology as well, as no prior information is used. For instance, it is possible to set an upper-limit of $A_{\rm ul}=0$ in a dataset without a (detectable) signal, which theoretically happens with probability $\alpha$.

It was shown in \citet{ccs+15} that the optimal statistic is identical to the cross-correlation statistic used by \citetalias{dfg+13}. This alternative interpretation of the optimal statistic allows us to obtain a measure of the cross power between pulsars. The cross power is the amount of correlated power between the timing residuals of different pulsars, and one would expect this cross power to follow the Hellings and Downs cross-correlation signature for a detectable GWB signal. The cross power and the uncertainty estimates are given by
\begin{equation}
\xi_{ab} = \frac{\delta\mathbf{t}_a^{\rm T}\mathbf{P}_a^{-1}\hat{\mathbf{S}}_{ab}\mathbf{P}_b^{-1}\delta\mathbf{t}_b}{{\rm tr}\left[\mathbf{P}_a^{-1}\hat{\mathbf{S}}_{ab}\mathbf{P}_b^{-1}\hat{\mathbf{S}}_{ba}\right]},
\quad
\sigma_{0,ab} = \left({\rm tr}\left[\mathbf{P}_a^{-1}\hat{\mathbf{S}}_{ab}\mathbf{P}_b^{-1}\hat{\mathbf{S}}_{ba}\right]\right)^{-1/2},
\label{eq:crosspower}
\end{equation}
which is independent of any specific overlap reduction function and $\hat{A}_{\rm gw}^2\Gamma_{ab}\hat{\mathbf{S}}_{ab} = \mathbf{S}_{ab}$. One can then fit these correlation coefficients assuming a particular overlap reduction function by minimizing
\be
\chi^{2} = \sum_{ab}\left| \frac{\xi_{ab}-A_{\rm gw}^2\Gamma_{ab}}{\sigma_{0,ab}}\right|^{2},
\ee
where $\Gamma_{ab}$ are the cross-correlation coefficients given by the overlap reduction function for an isotropic GWB. It can easily be shown that the best fit value of $A_{\rm gw}$ is then the optimal statistic value of Eq. \eqref{eq:os}.

\section{Results}
\label{sec:results}
\subsection{Optimal Statistic Analysis}
\label{sec:frequentist-analysis}

The optimal cross correlation statistic was applied to the full 9-year NANOGrav dataset. The maximum-likelihood single-pulsar noise values were obtained by independent noise analyses on each pulsar. The maximum-likelihood amplitude and SNR of this search are $\hat{A}_{\rm gw}=8.9\times 10^{-16}$ and $\rho=0.8$, respectively, indicating little evidence for the expected GWB cross correlations. The resulting upper limit using this method is $A_{\rm gw}^{95\%}=1.3\times 10^{{-15}}$ at a reference frequency of yr$^{-1}$, which is 5.4 and 1.5 times more stringent than the limits using this method presented in \citetalias{dfg+13} and \citetalias{ltm+15}, respectively. The corresponding signal-to-noise ratio is 0.8, indicating little evidence for the expected cross correlations from a GWB. It should be noted that the limiting technique presented in \cite{abc+09} and \citetalias{dfg+13} does not strictly have proper frequentist coverage in the presence of any measurable GWB signal\footnote{In the limit of zero GWB signal, this method actually over covers in the frequentist sense. We have found by comparison with other frequentist bounding techniques and Bayesian upper limits that the 5-year limit published in \citetalias{dfg+13} is indeed robust and does not suffer to under coverage.}; therefore, this limit will serve as a proxy to our improved sensitivity and not a strict upper limit.

\begin{figure}
\includegraphics[width=\columnwidth]{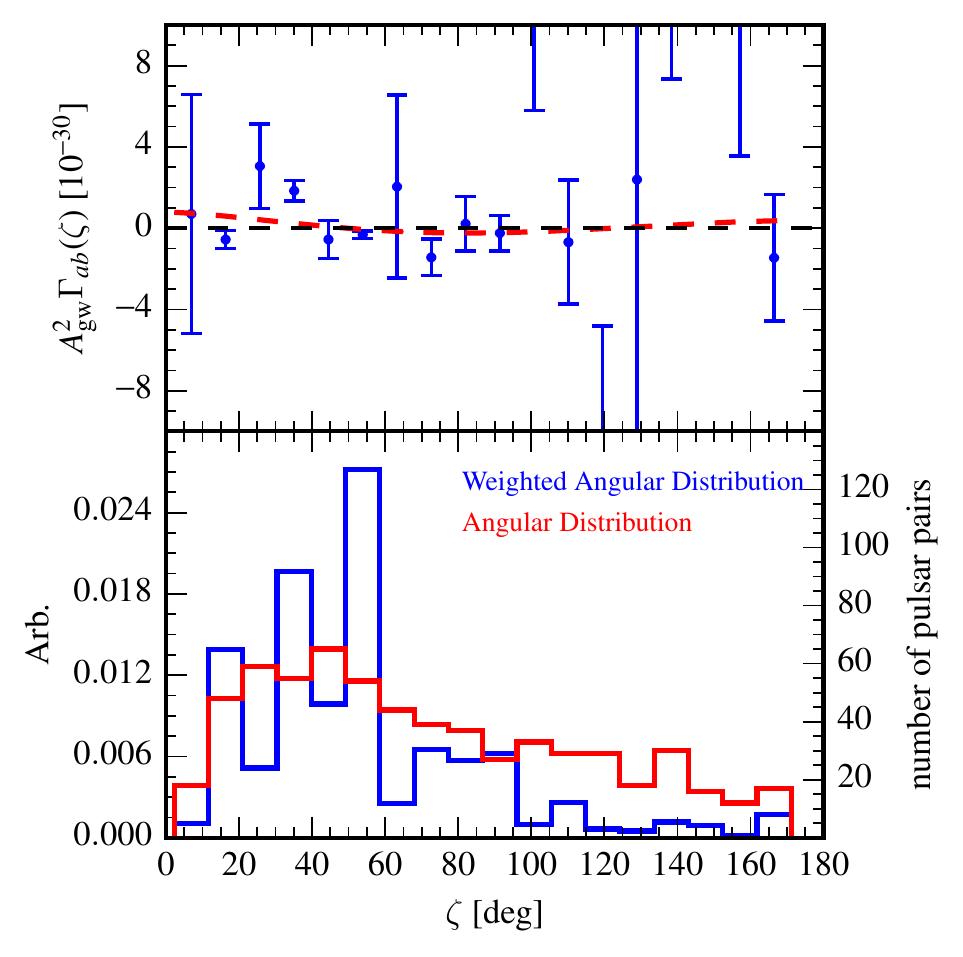} \\ 
\caption{(Top Panel): Cross-correlated power vs. angular separation obtained through an Optimal Statistic Analysis. The dashed red curve shows the maximum-likelihood amplitude mapped onto the Hellings and Downs coefficients. (Bottom Panel): Histogram of the number of pulsars in each bin (red right axis) and a weighted histogram (blue left axis) using the 1-$\sigma$ uncertainties in the cross correlation as weights. }
\label{fig:xcorr} 
\end{figure}

As a by-product of the optimal statistic analysis, we can obtain the maximum-likelihood cross-correlation values for each pulsar pair in the analysis. In Figure \ref{fig:xcorr} we plot the cross-correlated power vs. angular separation in the top panel. 
We have binned the values into 10 degree bins using a weighted averaging technique. The red curve shows the maximum-likelihood correlated power fit. It is clear that the cross-correlated power is still consistent with zero signal and that we have much more sensitivity at  values of $\zeta<100^{{\circ}}$, which is a by-product of the fact that our best timed pulsars all lie at a similar position on the sky. This is illustrated in the bottom panel of Figure \ref{fig:xcorr}, where we plot the histogram of the number of pulsars in each ten-degree bin (red) as well as the most significant bins (blue). To calculate the most significant bins we use the one-sigma uncertainties on each maximum-likelihood cross-correlation value as weights in the histogram. 

\subsection{Bayesian Analysis}
\label{sec:bayesian-analysis}

We have used the Bayesian data analysis techniques described in Section \ref{sec:data_analysis} to place upper limits on the GWB parameterized using the standard power law, spectral components and a broken-power-law. Bayesian data analysis of modern PTA data sets is quite difficult to perform in general due to the large parameter spaces ($\sim$ 500 for the current NANOGrav dataset assuming a power law red noise and GWB spectrum) and likelihood evaluation time. Even with the efficient methods described in Section \ref{sec:data_analysis} we are still prohibited from carrying out a full Bayesian search where we vary all noise parameters and GW parameters for all pulsars simultaneously. 

In this work we ameliorate these two problems by using a two-tiered approach to noise modeling and GWB analysis. To avoid including all noise parameters in our GWB analysis, we first carry out single-pulsar noise analyses on each pulsar in which we include EFAC, EQUAD, and ECORR for each backend/receiver system and red noise parameterized as a power law. We then perform the GWB analysis while holding all EFAC, EQUAD, and ECORR values fixed to their mean value from the single-pulsar analysis but allowing the red noise and GWB parameters to vary. By holding these white noise parameters fixed, we reduce the computational burden since large matrix products can be pre-computed. Furthermore, by holding all white noise values fixed we reduce the number of free parameters drastically from $\sim$ 500 to $2N_{\rm psr} + N_{\rm gw}$, where $N_{\rm psr}$ and $N_{\rm gw}$ are the number of pulsars in the array and number of parameters describing the GWB, respectively. 

We further reduce the computational burden in two ways: by ignoring cross correlations and by only using a subset of pulsars. We choose to ignore cross correlations in this work as they do not contribute to the upper limit in the absence of a signal and only serve to greatly increase our computational burden by requiring that we invert a large dense matrix for every iteration in the analysis. That the cross correlations have no bearing on the upper limit has been argued in \cite{src+13} and further shown in \citetalias{ltm+15}. A real GWB signal will reveal itself as a strong \emph{common} red noise term well before the cross correlations become detectable, and since we see no evidence for a common red noise term, we are justified in dropping these terms. Lastly, we choose to only include a subset of pulsars in our analysis as not all pulsars contribute to our upper limit either due to short timing baselines or large measurement errors. To choose this subset of pulsars, we first carry out our single pulsar analyses mentioned above but now include an extra red noise process with power spectral index fixed to $13/3$ (i.e., SMBHBs). The pulsars are then sorted based on their single pulsar upper limits on the GWB. We then carry out the GWB analysis mentioned above to compute an upper limit using an increasing number of pulsars in the sorted list. This process is continued until the upper limit saturates. In other words, including more pulsars beyond this point does not change the upper limit. Using this method we choose to use the 18 pulsars shown in Table \ref{tab:pulsars} .
\begin{table}
\begin{center}
\caption{\label{tab:pulsars} Pulsars used in GWB analysis.}
\begin{tabular}{lccc}
\hline\hline
PSR Name & $A_{\rm gw}^{95\%}$ $[10^{-15}]$ & RMS\footnote{Weighted root-mean-square of epoch-averaged post-fit timing residuals. See \cite{abb+15b} for more details.} $[\mu s]$ & Timing Baseline [yr] \\
\hline
J1713$+$0747 & 1.96 & 0.116 & 8.76 \\ 
J1909$-$3744 & 4.5 & 0.081 & 9.04 \\ 
J1640$+$2224 & 11.8 & 0.158 & 8.9 \\ 
J1600$-$3053 & 12.3 & 0.197 & 5.97 \\ 
J2317$+$1439 & 13.6 & 0.267 & 8.87 \\ 
J1918$-$0642 &  16.0 & 0.34 & 9.01 \\ 
J1744$-$1134 & 16.1 & 0.334 & 9.21 \\ 
J0030$+$0451 & 19.8 & 0.723 & 8.76 \\ 
J0613$-$0200 & 23.4 & 0.592 & 8.58 \\ 
J1614$-$2230 & 23.9 & 0.189 & 5.09 \\ 
B1855$+$09 & 26.6 & 1.338 & 8.86 \\ 
J1853$+$1303 &  31.0 & 0.235 & 5.6 \\ 
J2145$-$0750 &  33.0 & 0.37 & 9.07 \\ 
J1455$-$3330 & 37.9 & 0.694 & 9.21 \\ 
J1012$+$5307 & 43.3 & 1.197 & 9.21 \\ 
J1741$+$1351 & 56.8 & 0.103 & 4.24 \\ 
J2010$-$1323 & 83.3 & 0.312 & 4.08 \\ 
J1024$-$0719 & 92.9 & 0.28 & 4.01 \\ 
\hline
\end{tabular}

\end{center}
\end{table}

To compute the posterior probability of Eq. \eqref{eq:reducedlikelihood} and to map out the multi-dimensional parameter space we make use of the pulsar timing data analysis suite \textsc{PAL2}\footnote{\url{https://github.com/jellis18/PAL2}} in conjunction with the parallel-tempering Markov Chain Monte-Carlo (PTMCMC) code\footnote{\url{https://github.com/jellis18/PTMCMCSampler}}. The details of the PTMCMC sampler can be found in Appendix C of \cite{abb+14}. The parameters and prior ranges used in the analysis are shown in Table \ref{tab:pars}.
\begin{table*}[tp]
\begin{center}
\caption{\label{tab:pars} Summary model parameters and prior ranges.}

\begin{tabular}{lllc}
\hline\hline
Parameter & Description & Prior & Comments \\
\hline
\vspace{0.5em}
White Noise & & &\\
$E_{k}$ & EFAC per backend/receiver system & uniform in $[0, 10]$ & Only used in single pulsar analysis \\
$Q_{k}$ & EQUAD per backend/receiver system & uniform in logarithm $[-8.5, -5]$ & Only used in single pulsar analysis \\
$J_{k}$ & ECORR per backend/receiver system & uniform in logarithm $[-8.5, -5]$ & Only used in single pulsar analysis \\
\hline

\vspace{0.5em}
Red Noise & & &\\
$A_{\rm red}$ & Red noise power law amplitude & uniform in $[10^{-20}, 10^{-11}]$ & 1 parameter per pulsar \\
$\gamma_{\rm red}$ & Red noise power law spectral index & uniform in $[0, 7]$ & 1 parameter per pulsar \\
\hline

\vspace{0.5em}
GWB & & & \\
$A_{\rm gw}$ & GWB power law amplitude & uniform in $[10^{-18}, 10^{-11}]$ & 1 parameter for PTA for power-law models\\
$\gamma_{\rm gw}$ & GWB power law spectral index & delta function  & Fixed to different values depending on analysis\\
$\rho_{i}$ & GWB power spectrum coefficients at frequency $i/T$ & uniform in $\rho_{i}^{1/2}$ $[10^{-18},10^{-8}]^{a}$ & 1 parameter per frequency\\
$A$ & GWB broken power-law amplitude & log-normal$^{b}$ for models A(B)  & \\
& & $\mathcal{N}(-14.4(-15), 0.26(0.22))$ & 1 parameter for PTA for broken power law models\\
$\kappa$ & GWB broken power-law low-frequency spectral index & uniform in [0,7] & 1 parameter for PTA for broken power law models \\
$f_{\rm bend}$ & GWB broken power-law bend frequency & uniform in logarithm  [$-9$,$-7$]$^{c}$ & 1 parameter for PTA for broken power law models \\
\hline
\end{tabular}
\vspace{0.5em}
{$^{a}$ The prior uniform in $\rho_{i}^{1/2}$ is chosen to be consistent with a uniform prior in $A_{\rm gw}$ for the power law model since $\varphi_{i}\propto A_{\rm gw}^{2}$.}
\vspace{0.5em}

{$^{b}$ These values are quoted in log base 10 and are obtained from \citetalias{mop14} and \citetalias{s13}.}
\vspace{0.5em}

{$^{c}$ We choose different prior values on $\fbend$ when mapping to astrophysical model parameters as described in Section \ref{sec:discussion}.}

\end{center}
\end{table*}
As shown in the table, we have chosen to use uniform priors on both the red noise and GWB amplitude parameters. The GWB amplitude prior is chosen so that it is proper at $A_{\rm gw}=0$, that is to say, it must converge to a finite value in the limit of zero amplitude, otherwise the upper limit would depend on the lower bound of the prior which is undesirable. However, for the red noise amplitude we have no such restriction. A log-uniform prior would result in the most unbiased parameter estimation and a uniform prior would bias the parameter estimation towards higher red noise amplitude and lower red noise spectral indices (since there is a strong covariance between these two parameters); however, the log uniform prior would cause intrinsic red noise to be modeled by the GWB amplitude which is also undesirable. Because of these considerations we choose uniform priors on both as it gives equal weight to both red noise and the GWB. We also find that this prior results in a much more robust upper limit when using different numbers of frequencies in the rank-reduced approximations of Eq. \eqref{eq:expandedprior}.

\subsubsection{Power-law and Spectral Limits}
\label{sec:power-spectral-limits}

When computing upper limits on the dimensionless strain amplitude we fix the spectral index ($13/3$ in the case of SMBMBs) and adopt a uniform prior on $A_{\rm gw}$.  When performing a spectral analysis we again use priors that correspond to a uniform prior on $A_{\rm gw}$, this results in a prior that is uniform in the square root of the power spectrum coefficients. 
\begin{figure}
\centering
\includegraphics[scale=1]{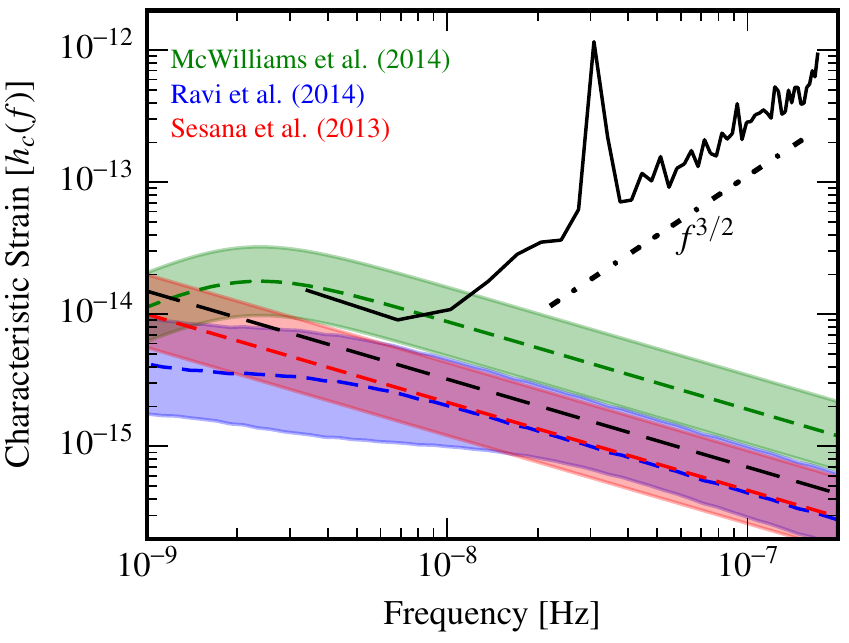} \\ 
\caption{Strain amplitude vs. GW frequency. The solid black and long dashed black lines are the 95\% upper limits from our spectral and power-law analyses. The red, blue and green shaded regions are the one-sigma predictions from the models of \citetalias{s13}, \citetalias{rws+14}, and \citetalias{mop14}. The green shaded region uses the simulation results from \citetalias{mop14}, but replaces the fit to the GWB predictions used in that paper with the functional form given by Eq. \eqref{eq:turnover}.}
\label{fig:ul-mcwilliams-ravi} 
\end{figure}

Figure \ref{fig:ul-mcwilliams-ravi} show the results of the power law and spectral analysis along with relevant astrophysical model predictions. The solid black and long dashed black lines are the 95\% upper limits from the spectral and power-law analyses, respectively. The green, blue, and red shaded regions are the one-sigma prediction on the strain spectra from \citetalias{mop14}, \citetalias{rws+14}, and \citetalias{s13}, respectively. We find an upper limit on the dimensionless strain amplitude of $A_{\rm gw}^{95\%}=1.5\times 10^{{-15}}$, a factor of 1.6 better than the most constraining published upper limit to date \citep{src+13}, a factor of 2 more constraining than the recent EPTA upper limit of \citetalias{ltm+15}, and a factor of 5 more constraining than the 5 year data release upper limit when applying the same Bayesian analysis. Furthermore, we find a slightly less constraining upper limit when using the free spectrum model (power-law equivalent upper limit of $2\times 10^{-15}$). This is to be expected since the free spectrum model has many more degrees of freedom (we use 50 free amplitudes for each of the 50 frequencies in this case) over the power law parameterization (1 degree of freedom). Thus, since the power law model can leverage extra information at all frequencies, as opposed to the spectrum model where each frequency is independent of the others, more constraining upper limits are expected from a power law model. We also find that the upper limit on the strain spectrum from the spectrum analysis is consistent with white noise (i.e., $h_{c}^{\rm white}(f)\propto f^{3/2}$) at frequencies $\gtrsim 3/T$, where $T$ is the length of the longest set of residuals in the data set, which indicates that our GWB upper limits are coming from the three lowest frequency bins. This behavior is to be expected since we have several well timed pulsars that do not span the full 9-year baseline (see Table \ref{tab:pulsars}) and thus will have peak sensitivity at frequencies greater than $1/T$. 

From inspection of Figure \ref{fig:ul-mcwilliams-ravi} we see that our 95\% upper limit is within at least the 2-sigma confidence region of all three astrophysical models and is sensitive to a potential turnover in the spectrum due to environmental coupling factors.  We wish to determine the level of consistency  between our data and the power-law models displayed in Figure \ref{fig:ul-mcwilliams-ravi}. To accomplish this we follow the method applied in \cite{src+13}. Given that we have a model $M$ for the value of the GW amplitude $A_{\rm gw}$ whose probability distribution function is denoted $p(A_{\rm gw}|M)$ and that we have a probability distribution function for $A_{\rm gw}$ given the data, denoted $p(A_{\rm gw}|d)$, where $d$ represents the data, the probability that we measure a value of $\hat{A}_{\rm gw}$ greater than that predicted by the model, $A_{\rm gw}^{M}$, is given by the law of total probablility
\be
\label{eq:totprob}
P(\hat{A}_{\rm gw} > A_{\rm gw}^{M}) = \int_{-\infty}^{\infty}p(A_{\rm gw}|M) dA_{\rm gw}\int_{A_{\rm gw}}^{\infty}p(A_{\rm gw}^{\prime}|d) dA_{\rm gw}^{\prime}.
\ee
Therefore, low values of $P(\hat{A}_{\rm gw} > A_{\rm gw}^{M})$ indicate that the range of $A_{\rm gw}$ that is consistent with our data is \emph{inconsistent} with the model $M$, and vice versa. To carry out this procedure the distribution $p(A_{\rm gw}|d)$ is simply the marginalized posterior distribution when using the uniform prior on $A_{\rm gw}$. We use log-normal distributions to model the \citetalias{mop14}, \citetalias{s13}, and \citetalias{rws+14}, models. Since the models of \citetalias{rws+14}, and \citetalias{s13}  predict nearly the same GWB amplitude distribution (assuming a power-law only) we make no distinction between these two models. Furthermore, the model distributions on $A_{\rm gw}$, given by log-normal distributions have mean and standard deviations of $(-14.4, -15)$ and $(0.26, 0.22)$ for the \citetalias{mop14} (hereafter model A) and \citetalias{s13}/\citetalias{rws+14} (hereafter model B) models, respectively. Using the aforementioned distributions and Eq. \eqref{eq:totprob} we find that our data are  0.8\% and 20\% consistent with models A and B, respectively,  under the assumption of a power-law. This indicates that either the assumptions that go into these models are incorrect, our universe is a realization of the GWB that has an amplitude in the tail of the probability distributions mentioned above, or that environmental effects are depleting SMBHB sources at low frequencies making the power-law assumption faulty. The implications of this last point are discussed in Section \ref{sec:discussion}.

\begin{figure}
\centering
\includegraphics[scale=1]{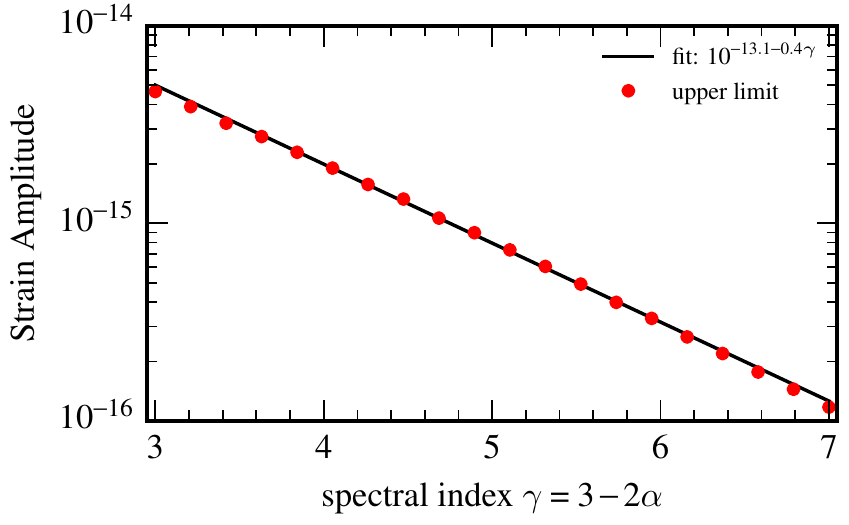} 
\caption{Upper limit on the GWB as a function of power spectral index.}
\label{fig:all-upper} 
\end{figure}

In addition to our power law limits on the stochastic background (i.e., strain spectral index $-2/3$), we have also computed the upper limit on a the GWB for a range of different spectral indices. In Figure \ref{fig:all-upper} we plot the upper limits obtained at varying spectral indices (red points) vs. power spectral index. We also provide the best fit model for the upper limit as a function of power spectral index where we find $A_{\rm gw}^{95\%}\propto10^{{-0.4\gamma}}\propto T^{0.83\alpha}$. This differs from \citetalias{dfg+13} where they find $A_{\rm gw}^{95\%}\propto T^{\alpha}$, arguing that this is due to the fact that the sensitivity is dominated by the lowest frequency of $1/T$. Our fit, giving a slightly weaker dependence on $\alpha$ is consistent with what we have seen above, namely that our limits are not completely dominated by the lowest frequency.

In a Bayesian analysis, the posterior distribution is the prior distribution updated by the data. Here we illustrate this by comparing our power-law upper limits, using identical methods, on the 5-year \citepalias{dfg+13} and 9-year \citep{abb+15b} NANOGrav data releases.
\begin{figure}
\centering
\includegraphics[scale=1]{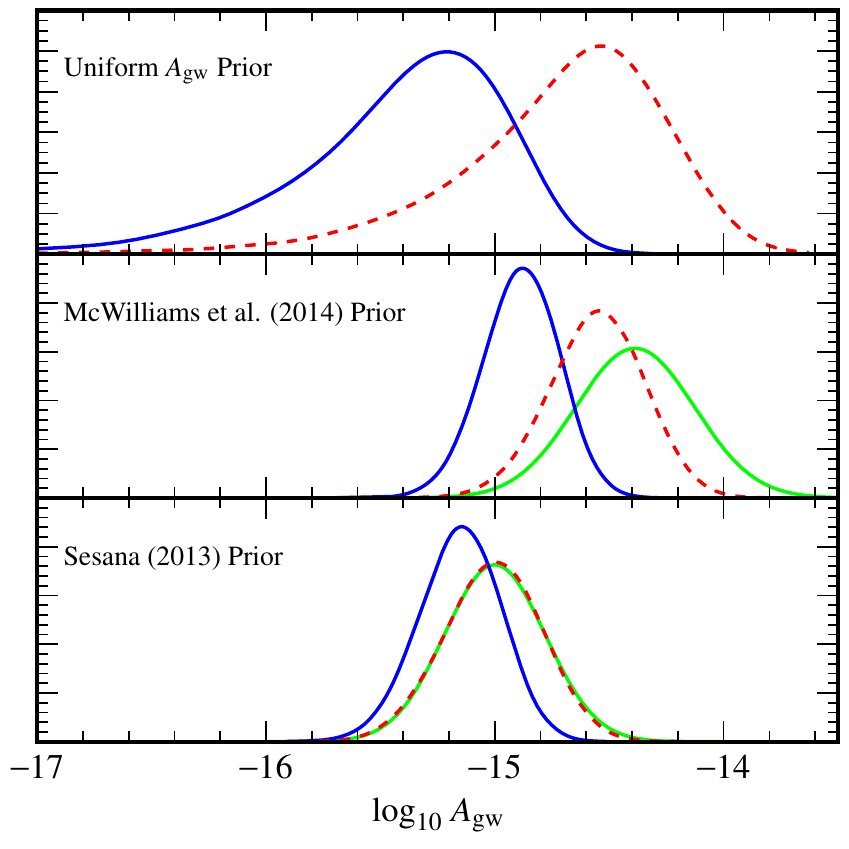} \\ 
\caption{Marginalized posterior probability density of $\log_{10} A_{\rm gw}$ computed using the nine (blue) and five (red) year NANOGrav data releases for uniform, \citetalias{mop14} model gaussian, and \citetalias{s13}/\citetalias{rws+14} model gaussian prior distributions. The gaussian priors are shown in green.}
\label{fig:9-5-amplitude} 
\end{figure} 
The results of this comparison are shown in Figure \ref{fig:9-5-amplitude} where we plot the marginalized posterior distributions of $\log_{10}A_{\rm gw}$ for the 9- and 5-year data releases in blue and red, respectively. The gaussian prior distributions described above are shown in green for model A and model B. For the uniform prior case we see quite a dramatic improvement (i.e., the factor of 5 mentioned above) in the upper limits.  For model A; the 5-year dataset does somewhat inform the prior, whereas the 9-year data set results in a posterior that is largely inconsistent with the prior distribution. For model B the 5-year data set do not inform the prior at all, whereas the 9-year data set does indeed update the prior.

\subsubsection{Broken Power-law Limits}
\label{sec:AstrophysicalModel}

We place constraints on the strength of environmental coupling effects that will likely affect our GWB signal at low frequencies (i.e., large orbital separations) via a simple parameterization of the GWB spectrum that allows for a ``bend'' frequency at which there is a transition from environmentally-driven evolution to GW-driven evolution. The following discussion and analysis techniques  are based on \cite{scm15}. Here we give a brief overview of this more generalized GWB spectrum.

The characteristic amplitude of a stochastic background from an ensemble of SMBHBs in circular orbits is \citep{p01, svc08, mop14}
\be
\label{eq:strain}
h_{c}(f)^{2} = \int_{0}^{\infty}dz \int_{0}^{\infty} d\mathcal{M} \frac{d^{3}N}{dz \,d\mathcal{M} \,dt}\frac{dt}{d \ln f} h^{2}(f),
\ee
where $d^{3}N/(dz \,d\mathcal{M} \,dt)$ is the differential number of inspiraling binaries per unit $z$, $\mathcal{M}$ and $t$, where $z$ is the redshift, $\mathcal{M}=(m_{1}m_{2})^{3/5}/(m_{1}+m_{2})^{1/5}$ is the chirp mass of the binary, and $t$ is the time measured in the binary rest frame. The $dt/d\ln f$ term describes the frequency evolution of the binary system, and $h(f)$ is the strain spectrum emitted by a single circular binary with orbital frequency $f/2$. Typically, it has been assumed that the binary is purely GW-driven which results in our usual expression for $h_{c}(f)$ given in Eq. \eqref{eq:gwstrain}; however, physical mechanisms other than GW radiation that are necessary to drive the binary to coalescence \citep{mm03} will change the frequency dependence (i.e., the $dt/d\ln f$ term) of this equation \citep[see][for a review of SMBHB coalescence]{c14}. Following \cite{scm15} we can generalize the frequency dependence of the strain spectrum to
\be
\label{eq:evolve-sum}
\frac{dt}{d\ln f} = f \lp\frac{df}{dt}\rp^{-1} = f\lp\sum_{i}\lp \frac{df}{dt} \rp_{i}\rp ^{-1},
\ee
where $i$ ranges over many physical processes that are driving the binary to coalescence. If we restrict this sum to GW-driven evolution and an unspecified physical process then the strain spectrum is now
\be
\label{eq:turnover}
h_{c}(f) = A\frac{(f/f_{\rm yr})^{\alpha}}{\lp 1+(\fbend/f)^{\kappa} \rp^{1/2}},
\ee
where $\fbend$ and $\kappa$ are the parameters that encode information about the physical processes (other than GW radiation) driving the binary evolution. As mentioned above, there could be many physical processes contributing to the frequency evolution of the SMBHB system; however, at current sensitivity it is very unlikely that our data can distinguish them. Thus we follow \cite{scm15} and adopt this slightly simplified spectrum to place constraints on possible environmental coupling mechanisms.

The above discussion has focused on the frequency evolution of SMBHB. The other piece of the equation is the merger rate of SMBHs, which will set the overall amplitude scale for the  GWB somewhat independently (assuming the last parsec problem is solved) of the physical mechanisms that drive the system to merger. Here we use the same log-normal distributions introduced in Section \ref{sec:power-spectral-limits} for models A and B as prior probability distributions for the GWB amplitude $A$ in Eq. \eqref{eq:turnover}.  In the following analysis we also fix $\alpha=-2/3$ and use uniform priors on $\log_{10} \fbend \in [-9, -7]$ and $\kappa \in [0,7]$ unless stated otherwise.

\begin{figure*}
\begin{center}
{\includegraphics[width=0.48\textwidth]{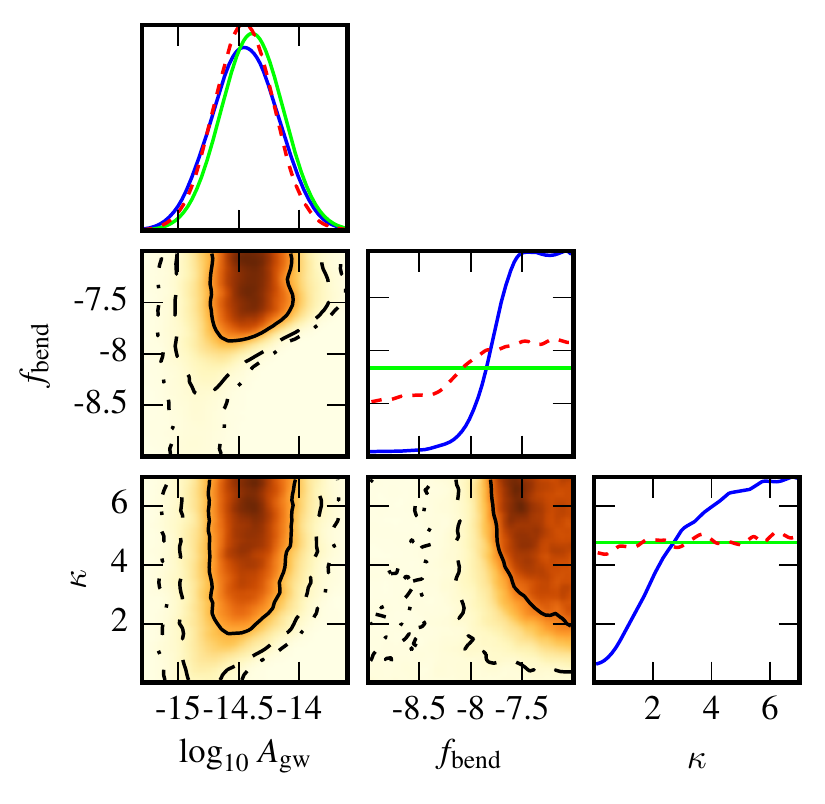}} \quad\quad
{\includegraphics[width=0.48\textwidth]{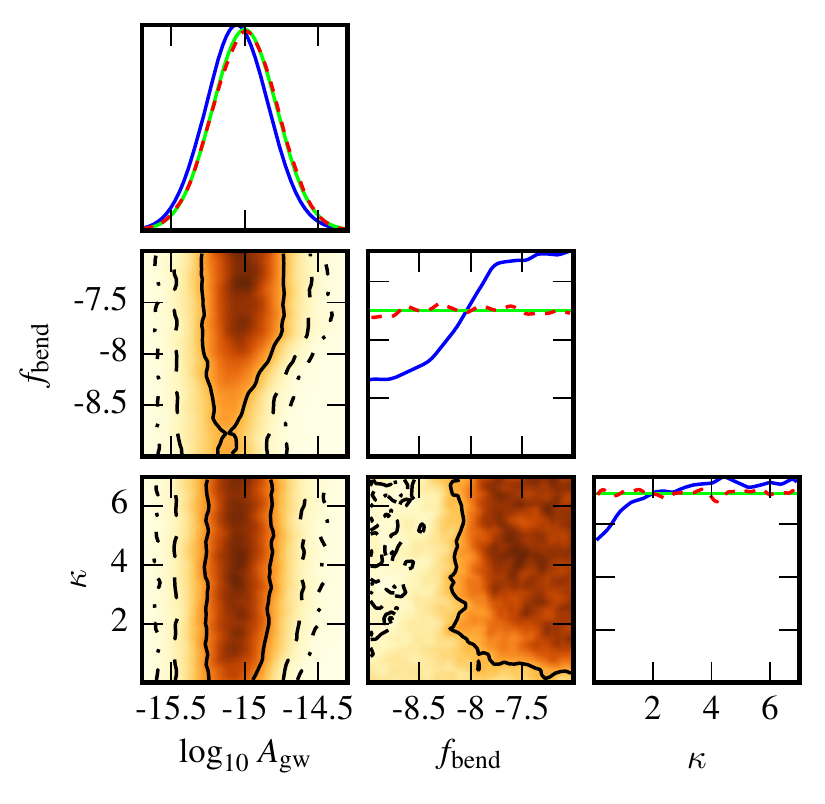}}
\end{center}
\caption{One- and two-dimensional posterior probability density plots of the spectrum model parameters $A_{\rm gw}$, $\fbend$, and $\kappa$. In the one-dimensional plots, we show the posterior probability from the 9-year data set (blue), the 5-year dataset (dashed red) and the prior distribution used in both analyses (green). In the two dimensional plots we show a heat map along with the one (solid), two (dashed), and three (dash-dotted) sigma credible regions. model A is on the left and model B is on the right.}
\label{fig:triplot}
\end{figure*}

Here, we have run an identical analysis to that of Section \ref{sec:power-spectral-limits} except that the GWB spectrum model is now that of Equation \eqref{eq:turnover} and we adopt the aforementioned priors on $A$, $\fbend$, and $\kappa$. 
\begin{figure*}
\begin{center}
\includegraphics[scale=1]{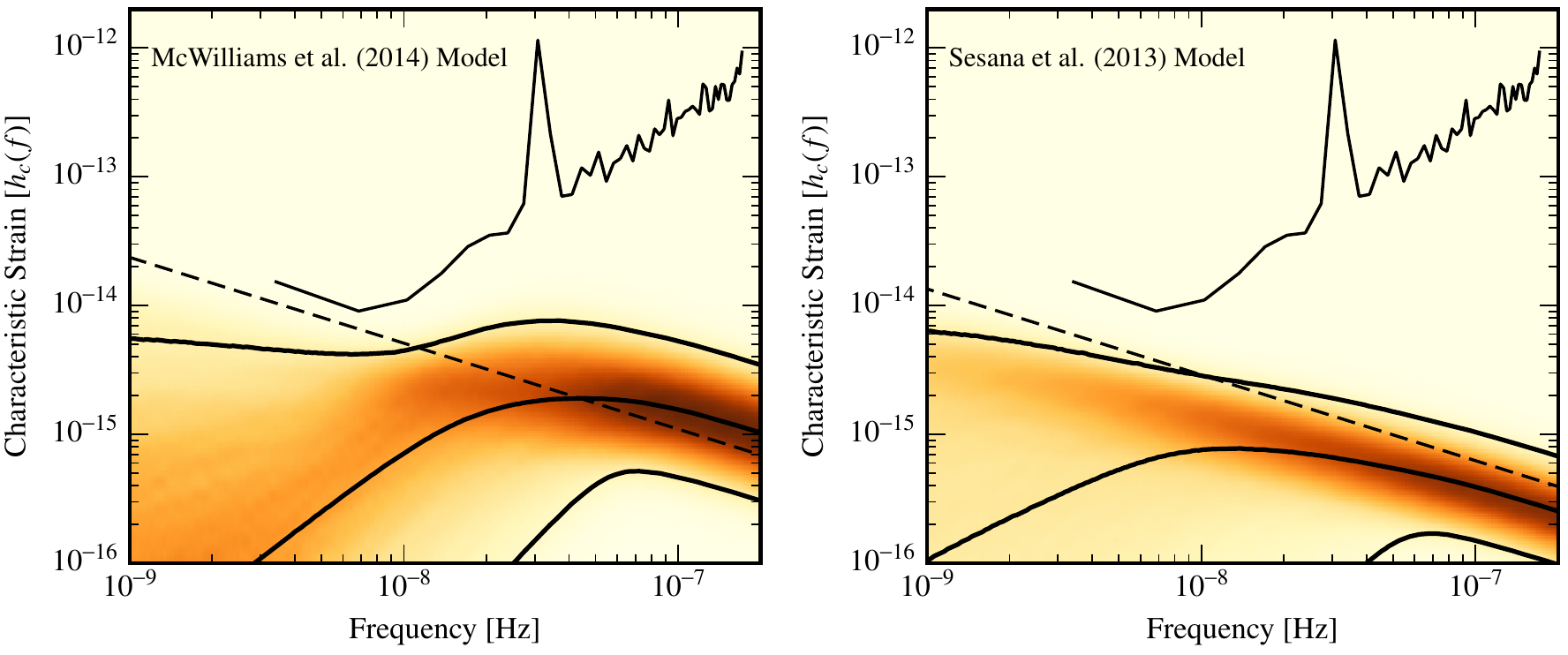}

\end{center}
\caption{Probability density plots of the recovered GWB spectra for models A and B using the broken-power-law model parameterized by ($A_{\rm gw}$, $\fbend$, and $\kappa$) as discussed in the text. The thick black lines indicate the 95\% credible region and median of the GWB spectrum. The dashed line shows the 95\% upper limit on the amplitude of purely GW-driven spectrum using the Gaussian priors on the amplitude from models A and B, respectively. The thin black curve shows the 95\% upper limit on the GWB spectrum from the spectral analysis.}
\label{fig:bayesogram}
\end{figure*}
Figures \ref{fig:bayesogram} and \ref{fig:triplot} show the results of this analysis. Figure \ref{fig:bayesogram} shows the posterior probability density of the GWB spectrum defined in Equation \eqref{eq:turnover} with model Aon the left and model B on the right. This probability density is constructed by drawing values of $A$, $\fbend$, and $\kappa$ from the joint probability distribution shown in Figure \ref{fig:triplot}, constructing the spectrum at each frequency via Equation \eqref{eq:turnover} and then producing a histogram of the spectral power at each frequency. The solid black lines in Figure \ref{fig:bayesogram} represent the 95\% credible region and the median of the GWB spectrum. The dashed line is the upper limit on $A_{\rm gw}$ using the purely GW-driven spectrum (i.e., no transition frequency) and the gaussian amplitude priors from models A and B, respectively. Lastly the thin solid black line is the 95\% upper limit on the GWB spectrum from the spectral analysis presented in Section \ref{sec:power-spectral-limits} . 
By inspecting the inferred GWB spectrum one can determine that the data prefer a GWB spectrum that has a definitive transition from GW-driven to environmentally-driven within the pulsar timing frequency band for Model A, whereas the data does not significantly constrain the shape of the spectrum for model B.

This can be seen further in the joint posterior distributions in Figure \ref{fig:triplot} in which the probability distributions (blue) for the bend frequency parameter, $\fbend$, and spectral index parameter, $\kappa$, are significantly different from the prior distribution (green) for model A and not significantly different for model B. When this same analysis is carried out on the 5-year data release we find that the data can only slightly inform the prior on $\fbend$ for model A and gives no information on the other parameters in either model A or B. This, again, indicates that the 9-year data release provides us with much more information about the shape of the GWB strain spectrum.

 Finally, we can be more quantitative and apply Bayesian model selection to this problem by computing the Bayes factor between the  broken-power-law (Equation \eqref{eq:turnover}) and pure-power-law (Equation \eqref{eq:gwstrain}) parameterizations for both models A and B. When this analysis is carried out we arrive at Bayes factors of $22.2 \pm 1.1$ and $2.23 \pm 0.15$ for models A and B, respectively. These Bayes factors were computed using parallel tempering and a custom thermodynamic integration implementation \citep[See Sec. 6.1.2 of][]{cl15}.

This analysis shows, for the first time, that PTAs are entering a regime where even in the case of a non-detection meaningful constraints can be placed on the dynamical history of the SMBHB population. Furthermore, this analysis shows that when placing upper limits to make statements about the full range of astrophysical merger scenarios it is no longer valid to consider only the classic strain amplitude, but one must instead frame the question in terms of measuring the amplitude \emph{and shape} of the GWB spectrum. As we have seen in the above analysis and as can be seen clearly in Figure \ref{fig:triplot}, given a model for the SMBHB merger physics (i.e., a prior on $A$) and discarding the assumption of a purely GW-driven signal (i.e., a broken-power-law model), it is very difficult to rule out any of the GWB amplitude parameter space with any certainty unless one has strong a-priori knowledge on the shape of the spectrum. However, we can begin to place constraints on the environmental coupling effects that drive the system to the GW-dominated regime.

\section{Discussion}
\label{sec:discussion}

\subsection{Astrophysical Model Limits}
While the parameter estimation of the previous section is interesting in and of itself, some of the most interesting science available from the NANOGrav data is accessible only by relating these parameters to properties of the source populations. Here we attempt to interpret the phenomenological posteriors on the GW spectrum  in terms of black hole-host galaxy relations, environmental effects, or binary-eccentricity by carrying out our analyses in sequence, investigating different effects \emph{separately}. In Section \ref{sec:joesarah} we use the results of our power-law analyses of Section \ref{sec:power-spectral-limits} to provide constraints on the parameters of scaling relations between host galaxies and black holes. We then go beyond the assumption of a power-law spectrum in Section \ref{sec:environs} to investigate how our constraints on the shape of the characteristic strain spectrum from Section \ref{sec:AstrophysicalModel} map to constraints on the environment of SMBH binaries. Finally in Section \ref{sec:eccentricity}, we probe the eccentricity of binaries before they entered the PTA band. 
\subsubsection{Constraints on Host Galaxy -- Black Hole Scaling Relations}
\label{sec:joesarah}
If the gravitational wave spectrum is assumed to be created by an ensemble of binary SMBHs that are formed following galaxy mergers (spectral index of $-2/3$), and whose evolution is assumed to be dominated by GW emission throughout the PTA-sensitive waveband, then we can trace the expected binary SMBH population using observations of galaxy merger rates, the galaxy stellar mass function, and the black hole-host galaxy relation. This is the approach taken in \citetalias{s13}, and \cite{rws+15}. Assuming equal fractional uncertainties in these parameters, the black hole-host galaxy relation will have the largest impact on the predicted level of the GW background. This is due to the much stronger dependence of the GW background on the chirp mass of each source than on the number of sources.

\cite{mdf+15} shows that it is difficult to extract information from PTA limits without making significant assumptions about the shape of the black hole merger rate density. If instead a galaxy merger rate density calculated from observed galaxy parameters is used as a proxy for the black hole merger rate density, then a limit on the GW background can be directly translated into a limit on the input galaxy parameter spaces \citep{ss15}. For this paper, we focus specifically on the scaling relation between host galaxy properties and black hole mass (e.\,g.\,~\cite{hr04}, \cite{mm13}) as it is the observed parameter that is most easily constrained by NANOGrav data. Specifically, we constrain the $M_{\bullet}-M_{\rm bulge}$ relation:
\be
\label{eq:M-M_bulge}
{\rm log}_{10} M_{\bullet} = \alpha + \beta ~{\rm log}_{10} \left( M_{bulge} / 10^{11} M_{\odot} \right)~.
\ee
In addition to $\alpha$ and $\beta$, observational measurements of this relation also fit for $\epsilon$, the intrinsic scatter of individual galaxy measurements around the common $\alpha$, $\beta$ trend line.  In practice, $\alpha$ and $\epsilon$ have the greatest impact on predictions of $A_{\rm gw}$, and all observational measurements agree with $\beta \approx 1$.

PTAs are most sensitive to binary SMBHs where both black holes are $\gtrsim$10$^8\msun$ (e.g. \cite{svc08}). Therefore $M_{\bullet}-M_{bulge}$ relations that are derived including the most massive systems are the most relevant to understanding the population in the PTA band. Several recent measurements of the $M_{\bullet}-M_{bulge}$ relation specifically include high-galaxy-mass measurements, e.g. those from Brightest Cluster Galaxies (BCGs). As these fits include the high-mass black holes that we expect to dominate the PTA signals, we take these as the ``gold standard" for comparison with PTA limits \citep[e.g.]{kh13, mm13}.

\begin{figure}
\centering
\includegraphics[width=\columnwidth]{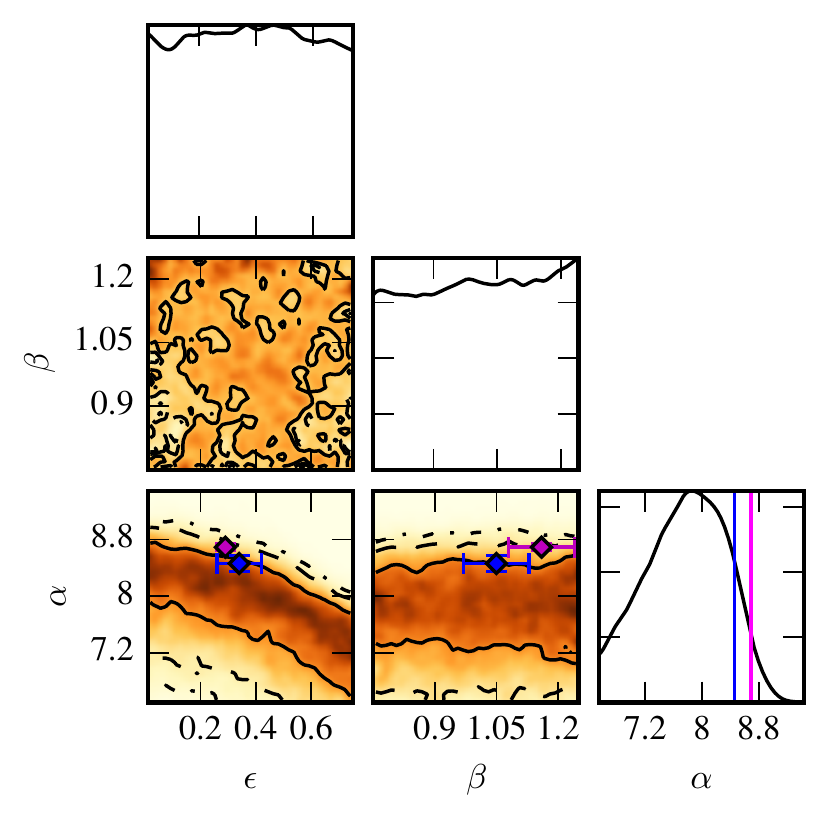} 
\caption{The above plot shows the translation of the marginalized posterior distribution of $A_{\rm gw}$, Fig. \ref{fig:9-5-amplitude}, into the black hole-host galaxy parameter space, which is characterized by an intercept $\alpha$, a slope $\beta$, and an intrinsic scatter $\epsilon$. Flat priors are used for $\alpha$, $\beta$, and $\epsilon$. $\beta$ is not informed by the distribution of $A_{\rm gw}$, while both $\alpha$ and $\epsilon$ are, with a limit on $\alpha$ being more strongly set. The curves show the 1, 2, and 3$\sigma$ contours. Relevant observational measurements are also shown, with \cite{mm13} in blue and \cite{kh13} in magenta. Since $\beta$ is not strongly informed by the upper limit, we can set an upper limit in $\alpha$-$\epsilon$ space by marginalizing over $\beta$. That upper limit is shown in Fig. \ref{fig:alpha-epsilon}.}
\label{fig:a-b-e-triplot}
\end{figure}

The translation of an upper limit on $A_{\rm gw}$ to the black hole-host galaxy parameter space is calculated as follows: 
\be
p \left( \alpha, \beta, \epsilon | {\rm PTA} \right) \propto \int d\theta ~p(\theta) ~p(A_{\rm gw} ( \alpha, \beta, \epsilon, \theta ) | {\rm PTA}) ,
\ee
where the posterior of $A_{\rm gw}$, $p \left( A_{\rm gw} (\alpha, \beta, \epsilon, \theta) | {\rm PTA} \right)$, is the marginalized posterior distribution of $A_{\rm gw}$, which is shown in Fig. \ref{fig:9-5-amplitude}; $A_{\rm gw} \left( \alpha, \beta, \epsilon, \theta \right)$ is the prediction of $A_{\rm gw}$ calculated from models similar to \citetalias{s13}; $\theta$ represents the galaxy stellar mass function and the galaxy merger rate; and $p \left( \alpha, \beta, \epsilon | {\rm PTA} \right)$ is the marginalized posterior distribution of the black hole-host galaxy relation, which is shown in Fig. \ref{fig:a-b-e-triplot}. For this analysis, we use two leading measurements of the galaxy stellar mass function, \cite{iml+13} and \cite{tqt+14}, and two measurements of the galaxy merger rate, \cite{rdd+14} and \cite{kfp+14}, as the basis for simulating a local population of binary SMBHs. A flat prior is used for $\alpha$, $\beta$, and $\epsilon$, and the posterior on $A_{\rm gw}$ using a uniform prior, as seen in Fig. \ref{fig:9-5-amplitude}, is directly translated into this parameter space. The result of which is shown in Fig. \ref{fig:a-b-e-triplot}. $\beta$ is clearly not informed by a PTA posterior, but the combination of $\alpha$ and $\epsilon$ are, with the strongest limit being set on $\alpha$.

\begin{figure}
\centering
\includegraphics[width=\columnwidth]{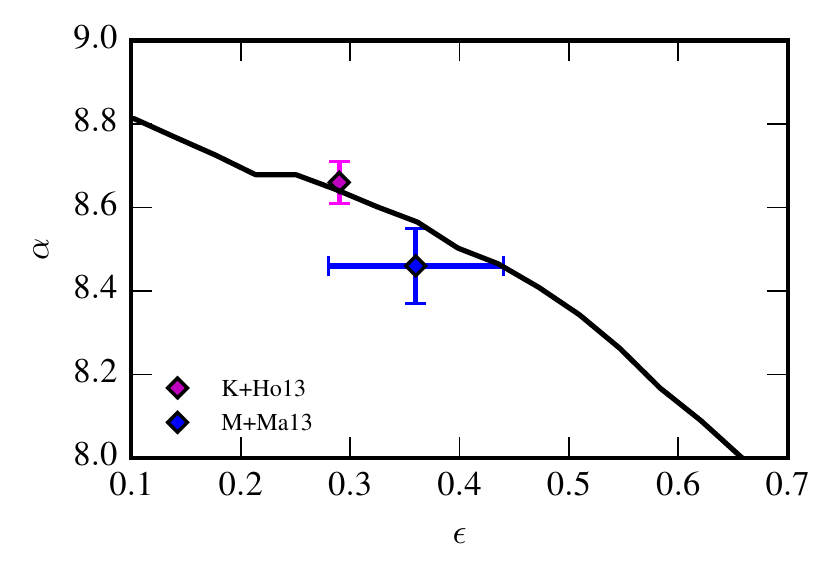} 
\caption{The above plot shows the translation of the $95\%$ upper limit on $A_{\rm gw}$, Fig. \ref{fig:9-5-amplitude}, into the parameter space $\alpha$-$\epsilon$, which characterizes the black hole-host galaxy relation as described in Equation. \eqref{eq:M-M_bulge}. The parameter space above the line is inconsistent with the power-law analysis of the \citetalias{s13} model, as described in \cite{ss15}. Observational measurements of this parameter space are shown with errorbars.}
\label{fig:alpha-epsilon}
\end{figure}

Fig. \ref{fig:alpha-epsilon} shows the translation of our posterior on $A_{\rm gw}$ into $\alpha$-$\epsilon$ parameter space with observational measurements of the parameters from \cite{kh13} and \cite{mm13}. Assuming a power-law analysis of the \citetalias{s13} model, as described in \cite{ss15}, there is a slight inconsistency between our upper limit and the \cite{kh13} measurement.
Potential solutions to an inconsistency include: a significant number of black hole binaries do not reach the GW-dominant regime in our assumed timescale (they ``stall"); the `classical' assumption of a power-law strain spectrum in the PTA band is incorrect and in fact there is a turn-over in the strain spectrum at lower frequencies (see Sec. \ref{sec:AstrophysicalModel}); or that the measured astronomical parameters are not correct for the population of binary SMBHs in the PTA band.

\begin{figure}
\centering
\includegraphics[width=\columnwidth]{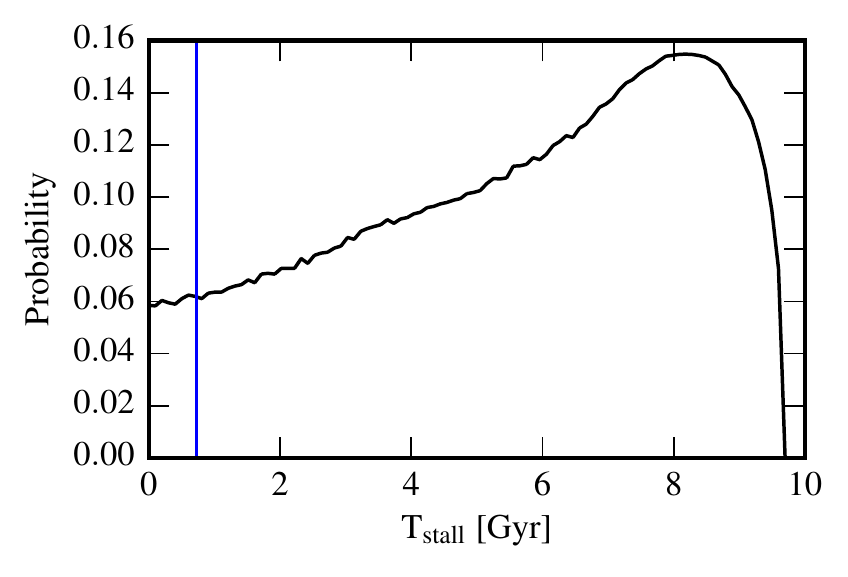} 
\caption{By introducing the parameter $T_{\rm stall}$, as described in \cite{ss15}, we can start to explore the inconsistency of our upper limit with power-law models for the GW background. In the above plot, we allow $T_{\rm stall}$ to vary while using the $M_{\bullet}-M_{\rm bulge}$ relation \cite{kh13}. The probability of $T_{\rm stall}$ is a direct translation of the posterior on $A_{\rm gw}$ from Fig. \ref{fig:9-5-amplitude}. The blue line is the $95\%$ lower limit on $T_{\rm stall}$, which we set at 0.73 Gyr. While this is not sufficiently constraining to make meaningful astrophysical statements, this parameter may be useful for future PTA upper limits.}
\label{fig:Tstall}
\end{figure}

As the possibility for a different strain spectrum curve is discussed in Sec. \ref{sec:AstrophysicalModel}, let us explore the potential for `stalling' within the model described so far in this section. Using the galaxy merger rate density as a proxy for the black hole merger rate density implies as assumption that the events occur at a similar cosmological time. If there was significant stalling in the binary black hole population, then these events would be offset in cosmological time by some `stalling timescale'. There is then nothing in the model to allow for anything other than an efficient solution to the `last-parsec' problem \cite{bbr80}.
As done in \cite{ss15}, we introduce a variable, $T_{\rm stall}$, which is a measure of this offset in time between the assumed galaxy merger rate density and the black hole merger rate density used to model a GW background. Fig. \ref{fig:Tstall} shows the translation of the posterior distribution of $A_{gw}$, Fig. \ref{fig:9-5-amplitude}, into a probability distribution of $T_{\rm stall}$ using the \cite{kh13} measurements of the $M_{\bullet}-M_{\rm bulge}$ relation. Using this we set a $95\%$ lower limit on $T_{\rm stall}$ of 0.73 Gyr, which is not sufficiently constraining to indicate which of the many suggested `solutions' to the `last-parsec' problem are not occurring for the systems in the PTA band. However, this parameter may be useful for future PTA upper limits as a probe of the level of `stalling' in the binary black hole population.

\subsubsection{Constraints on binary environmental influences}
\label{sec:environs}

The cores of galactic merger remnants can harbor stars with little angular momentum and almost radial trajectories which intersect the central galactic region (centrophilic orbits). These stars can undergo three-body interactions with the resident supermassive black-hole binary, causing the stars to be ejected, which results in energy and angular momentum being extracted from the black hole system, and leading to binary hardening \citep{q96}.\footnote{We assume that all galactic merger remnants maintain the same mass density of core stars throughout the binary merger. The subtleties of loss-cone replenishing impact the evolution of the binary and of the central density profile within a factor of $\sim 2$ (a few at most), as shown by \citet{sk15} and \citet{vam15}.} Additionally, the formation of circumbinary gaseous disks can lead to interactions which extract energy and angular momentum from the binary orbit, driving it towards smaller orbital separations \citep{ipp99}. We expect that, in the type of gas-poor galaxies which dominate the nanohertz GW signal, hardening from stellar scattering will dominate over circumbinary interactions, but we consider both in the following. We begin with a discussion of how these environmental mechanisms drive the evolution of the binary, then provide constraints on the frequency at which the characteristic strain spectrum exhibits a turnover from the analysis in Sec.\ \ref{sec:AstrophysicalModel}. We finish by mapping these frequencies to constraints on the astrophysical environment of SMBH binaries emitting GWs in the nanohertz band. 

\paragraph{Environmentally-driven orbital evolution}

We use the formalism of \citet{q96} to define the evolution of a (circular) binary due to three-body stellar scattering events, where
\be 
\frac{{\rm d}}{{\rm d}t}\left(\frac{1}{a}\right) = \frac{G\rho H}{\sigma},
\ee
where $a$ is the binary orbital semi-major axis, $\rho$ is the mass density of galactic core stars, $H$ is a dimensionless hardening rate which takes a value of $\sim 15$, and $\sigma$ is the velocity dispersion of core stars. Using Kepler's third law, we can rearrange this equation to solve for the rate of orbital frequency evolution, which gives ${\rm d}f/{\rm d}t \propto f^{1/3}$. Since the binary orbital evolution will be due to a combination of environmental influences and GW emission, ${\rm d}f/{\rm d}t$ is actually a sum over all mechanisms, as in Eq.\ (\ref{eq:evolve-sum}). We know that the rate of frequency evolution due to GW emission is ${\rm d}f/{\rm d}t \propto f^{11/3}$ \citep{pm63}, hence, in the language of the parametrized spectrum model in this paper given in Eq.\ (\ref{eq:turnover}), $\kappa=10/3$ for binary hardening by three-body stellar scattering. 

Likewise, the evolution of a circular binary due to circumbinary disk interaction is modeled within the $\alpha$-disk formalism \citep{ipp99,hkm09,s13c} as \citep{s13c}
\be
\frac{{\rm d}a}{{\rm d}t} = -\frac{2\dot{M}_1}{\mu}(aa_0)^{1/2},
\ee
where $\dot{M}_1$ is the accretion rate onto the primary black hole, $\mu$ is the binary reduced mass, and $a_0$ is a characteristic orbital separation at which the enclosed disk mass equals the mass of the secondary black hole. The latter can be expressed as \citep{ipp99}
\begin{align}
a_0 =&\,\, 3\times 10^3 \left(\frac{\alpha}{10^{-2}}\right)^{4/7}\left(\frac{M_2}{10^6 M_\odot}\right)^{5/7}\ \nonumber\\
&\times\left(\frac{M_1}{10^8 M_\odot}\right)^{-11/7}\left(100\frac{\dot{M_1}}{\dot{M_\mathrm{E}}}\right)^{-3/7}r_g,
\end{align}
where $\alpha$ is a disk viscosity parameter, $M_{1,2}$ are the binary black hole masses; $\dot{M}_\mathrm{E} = 4\pi GM_1/c\kappa_{\rm T}$ is the Eddington accretion rate of the primary ($\kappa_{\rm T}$ is the Thompson opacity coefficient); and $r_g=2GM_1/c^2$ is the Schwarzschild radius of the primary. 

As in the stellar scattering case, we can rearrange this equation to determine the orbital frequency evolution. In this model, ${\rm d}f/{\rm d}t\propto f^{4/3}$, and so $\kappa=7/3$ for $\alpha$-disk binary interactions.

\paragraph{Constraints on spectral turnover frequency}
We define the spectral turnover frequency in the obvious way to mean the frequency at which the characteristic strain spectrum exhibits a change in slope. If the low-frequency slope is positive, this will correspond to the point at which the spectrum is maximized. One must note that setting $f=f_{\rm bend}$ in Eq.\ (\ref{eq:turnover}) does not maximize the characteristic strain spectrum. Rather the turnover frequency will be a function of both $f_{\rm bend}$ and $\kappa$:
\be
\label{eq:max-freq}
f_{\rm turn} = f_{\rm bend}\left(\frac{3\kappa}{4}-1\right)^{1/\kappa}.
\ee
We can combine our measurements of $f_\mathrm{bend}$ and $\kappa$ from the analysis in Sec.\ \ref{sec:AstrophysicalModel} to compute the probability distribution of spectral turnover frequencies. We find that placing numerical constraints on $\fbend$ is difficult as the posterior is heavily influenced by the prior, namely the upper and lower bounds for the uniform priors used in this analysis. In the following analysis we set the lower bound on $\fbend$ by requiring that the power at $f=1/T_{\rm span}$ differs by no more than 10\% from a pure-power-law for any value of $\kappa$. By using this prior, we ensure that we can recover a pure-power law spectrum (in our frequency range) for any value of $\kappa$. The upper bound of $\fbend$ is chosen based on the specific environmental coupling mechanism we are considering.

We emphasize that the probability
distributions of $f_\mathrm{bend}$, $\kappa$, and $f_\mathrm{turn}$ are not distributions of these parameters over the SMBH binary population. Rather our posterior distribution is illustrating the spread of our beliefs in the measurement of the single $f_\mathrm{bend}$ and $\kappa$ model parameters. 

%
%

\paragraph{Constraints on environmental parameters}
We can now extract astrophysical constraints from our constraints on the transition and spectral turnover frequencies. By equating the rate of orbital evolution, ${\rm d}a/{\rm d}t$, due to environmental mechanisms and GW emission, we can deduce the characteristic transition frequency, $f_{\rm bend}$, between these influences. We firstly consider stellar-scattering, for which the transition frequency is given by

\begin{align}
\label{eq:fbend_stars}
f_\mathrm{bend} &= 3.13~\mathrm{nHz}\times\sigma^{-3/10}_{200} \rho^{3/10}_3 H^{3/10}_{15} M^{-2/5}_8 q^{-3/10}_r \\
&= 3.25~\mathrm{nHz}\times \rho_3^{3/10} H_{15}^{3/10}
M_8^{-23/50} q_r^{-3/10}
\end{align}
where $M$ is the total binary mass, $M_8 \equiv M / (10^8~\Msun)$, $q=M_2/M_1$, $q_r=q/(1+q)^2$, $\sigma_{200} \equiv
\sigma/(200~\mathrm{km/s})$, $\rho_3 \equiv \rho /
(10^3~\Msun\mathrm{pc}^{-3})$, and $H_{15} \equiv H / 15$. In the second line we have used the $M-\sigma$ relation \citep{mmg+11} to replace velocity
dispersion, where $M_8 \approx 1.9\;
\sigma_{200}^5$. It follows from the \citet{mmg+11} $M-\sigma$ relation that the velocity dispersion term on the first line of Eq.\ (\ref{eq:fbend_stars}) has the mass scaling $\sigma_{200}^{-3/10}\propto M_8^{-3/50}$, which is a small modification to the exponent of the other mass term, $M_8^{-2/5}$. Hence, $f_\mathrm{bend}$ has a relatively weak dependence on the mass scaling of $\sigma$. Nevertheless we can express the transition frequency in terms of variables of a parametrized $M-\sigma$ relation, where ${\log_{10}(M/M_\odot) = a + b\log_{10}\sigma_{200}}$, such that $M_8 = 10^{a-8}\sigma_{200}^b$, and 
\be
f_\mathrm{bend} = 3.13~\mathrm{nHz}\times 10^{3(a-8)/10b} \rho_3^{3/10} H_{15}^{3/10}
M_8^{-(3+4b)/10b} q_r^{-3/10}.
\ee
Finally, the weak scaling of $f_\mathrm{bend}$ with $H$, and the $\lesssim 20\%$ deviations of this parameter away from $15$ seen in numerical scattering experiments \citep{sk15} justifies our keeping this parameter fixed at its fiducial value of $H_{15}=1$. Astrophysical estimates on $\rho$ are quite uncertain with estimated values around $10-10^4~\Msun$~pc$^{-3}$~for typical environments~\citep{dch+07}. The variation of estimates over several orders of magnitude is why we choose to investigate only $\rho$ in our stellar-scattering constraints. 

The equivalent transition frequency for $\alpha$-disk interaction is
\be
\label{eq:fbend_gas}
f_\mathrm{bend} \approx 0.144~\mathrm{nHz}~M^{-17/14}q_r^{-6/7}\dot{M}_1^{3/7}a_0^{3/14},
\ee
where $a_0$ is as previously defined. We adopt the fiducial value of the disk viscosity parameter $\alpha\sim10^{-2}$ used in \citet{ipp99}. The very weak dependence of the bend frequency on this parameter, $f_\mathrm{bend}\propto \alpha^{6/49}$ will significantly dampen the influence of any deviations from this fixed value. Hence, in our constraints on the influence of circumbinary disk interactions, we only vary the accretion rate of gas onto the primary black-hole, $\dot{M}_1$, of which estimates in the literature vary over several orders of magnitude -- typically $10^{-3}\Msun$~yr$^{-1}$ -- $1~\Msun$~yr$^{-1}$, see e.g. \citealt{mcf01, an02, gcs+15}.

Equations (\ref{eq:fbend_stars})-(\ref{eq:fbend_gas}) indicate the GW frequency at which a \textit{single} circular binary will transition between being environmentally driven and being GW driven. Of course, $f_\mathrm{bend}$ is not the quantity that can be extracted from a spectral analysis -- $f_\mathrm{turn}$ is what we can measure. Our analysis of the stochastic GW background has provided us with constraints on the characteristic transition frequency of the entire population, and thus the turnover of the spectrum. Hence, our path to providing constraints on environmental parameters requires us to numerically construct characteristic strain spectra for populations of SMBH binaries in contact with their environment. We can then construct numerical mappings between the environmental parameters of interest (core stellar mass density, $\rho$, and primary black hole accretion rate, $\dot{M}_1$) and the turnover of the spectrum. We use the formalism of \citet{p01} via Eq.\ (\ref{eq:strain}), where the differential comoving number density of merging binaries per redshift and component masses is constructed as in \citet{mop14}:
 \be 
 \label{eq:merger-density}
 \frac{{\rm d}^3 n}{{\rm d} z{\rm d} M_1{\rm d}M_2} \propto P(z)\phi(M_1)\phi(M_2),
 \ee
 where $P(z)$ encapsulates redshift dependent factors, and $\phi(M_{\{1,2\}})$ is the number density of black holes of a certain mass, which is given by a (redshift dependent) Schechter function modified at the high-mass end by a lognormal component to accommodate recent high mass BCG (brightest cluster galaxy) discoveries \citep{lom10}. 
 
\begin{figure*}[!t]
\begin{center}
{\includegraphics[width=0.48\textwidth]{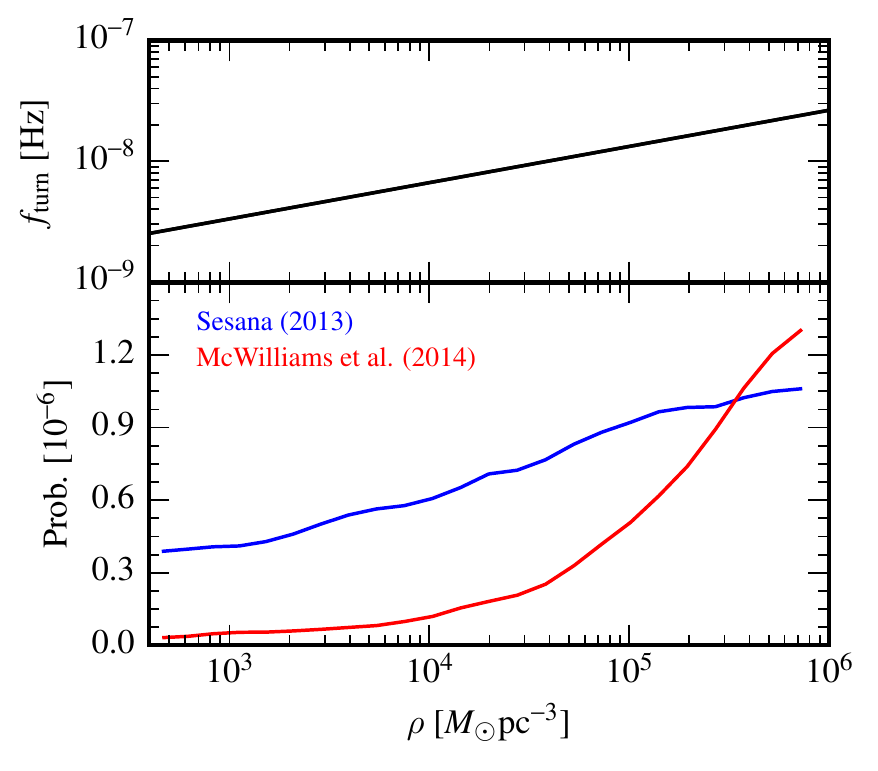}} \quad\quad
{\includegraphics[width=0.48\textwidth]{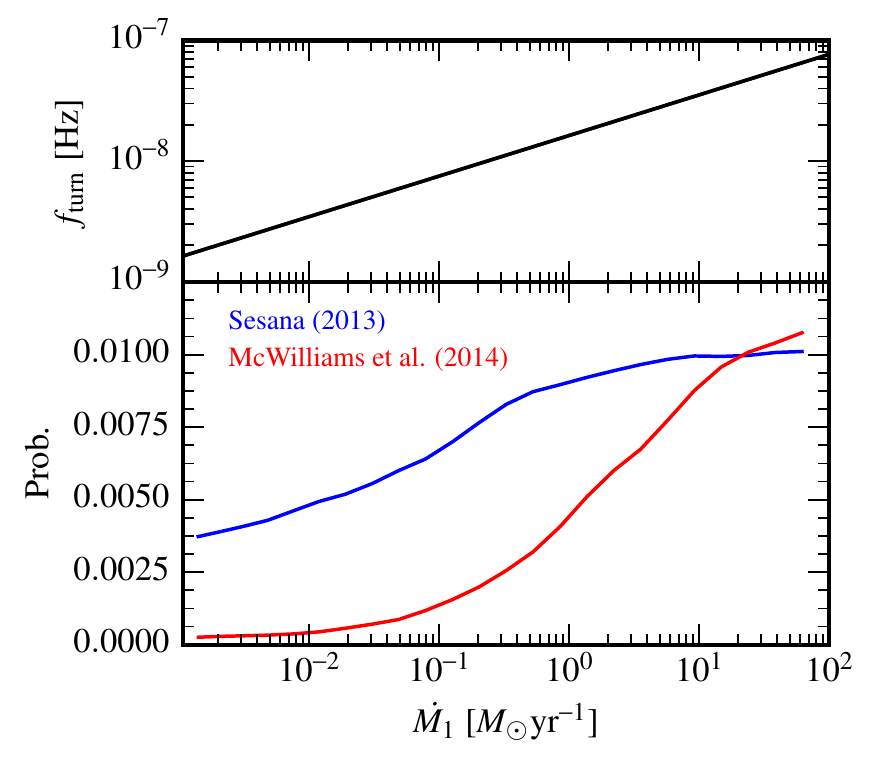}}
\end{center}
\caption{(\emph{top}): Empirical mapping from $f_{\rm turn}$ to $\rho$ (\emph{left}) and $\dot{M}_{1}$ (\emph{right}). (\emph{bottom}): Posterior distributions for the mass density of stars in the galactic core (\emph{left}) and the accretion rate of the primary black hole from a circumbinary disk (\emph{right}). These distributions are constructed by first converting the marginalized distribution of $\fbend$ to a distribution of $f_{\rm turn}$ via Eq. \eqref{eq:max-freq}, and then using the empirical mapping described in the text to convert from $f_{\rm turn}$ to the astrophysical quantities $\rho$ and $\dot{M}_{1}$, respectively.}

\label{fig:env_lims} 
\end{figure*} 
 The combined influence on the binary orbital evolution of GW emission and environmental couplings is modeled with the sum in Eq.\ (\ref{eq:evolve-sum}), where either stellar scattering \textit{or} disk interactions are included i.e. we consider one mechanism at a time. By considering all binary environments to have the same $\rho$ or $\dot{M}_1$, we can determine the spectral turnovers from the numerically computed strain spectra, iterating over many values of these environmental parameters to deduce a mapping. We draw binary systems from the ranges $z\in [0,1]$, $M_1\in [10^7, 10^{10}]~M_\odot$, and $q\in [0.1,1]$. The results for our fiducial assumptions are shown in the top panel of Fig.\ \ref{fig:env_lims}, with the stellar density required to give a certain turnover frequency shown in the left panel, and the primary accretion rate required to give a certain turnover frequency shown in the right panel.
  
In the lower panels of Fig. \ref{fig:env_lims} we plot the posterior distributions of the stellar density, $\rho$, for stellar hardening, and mass accretion rate, $\dot{M}_{1}$ for circumbinary disk interaction. In this analysis we perform the Bayesian parameter estimation for fixed values of $\kappa$ corresponding to the appropriate values for stellar hardening  ($\kappa=10/3$) and circumbinary disk interaction  ($\kappa=7/3$). These posteriors are constructed by first converting the marginalized distributions on $\fbend$ to a distribution for $f_{\rm turn}$ via Eq. \eqref{eq:max-freq} and then using the empirical mapping to convert $f_{\rm turn}$ to the appropriate astrophysical parameter. Again, we do not place numerical confidence limits on $\rho$ or $\dot{M}_{1}$ since the data does not constrain the prior distribution at large values. Nonetheless, from inspection of Fig. \ref{fig:env_lims} we see that the \citetalias{mop14} model heavily prefers $\rho\gtrsim10^{4}$ $M_{\odot}$pc$^{-3}$ and $\dot{M}_{1}\gtrsim10^{-1}$ $M_{\odot}$yr$^{-1}$, while the \citetalias{s13} model is largely unconstraining for both mechanisms.  
Typical densities of massive elliptical galaxies at the MBH influence radius is $\sim 10^{3}$ $\Msun$ pc$^{-3}$, making the \citetalias{mop14} model hard to reconcile with observations, even if we consider that massive ellipticals were likely factor 2--3 more compact at $z=1$ (Sesana, unpublished). Our results approach the upper end (for the \citetalias{mop14} prior) of the expected range of $\dot M$, $10^{-3}\Msun$~yr$^{-1}$ -- $1~\Msun$~yr$^{-1}$, observed in the local universe and predicted via simulations, see e.g. \citealt{mcf01, an02, gcs+15}. Furthermore, \cite{dmm15} predict that $\dot{M}_{1}\ll 10^{-1}$ $\Msun$~yr$^{-1}$ for BH masses of $10^{9}$ $\Msun$ and redshifts $z<1$; however, these are average accretion rates, and short, episodic accretion triggered by galaxy mergers could occur at a higher rate.


 
 We go beyond the fiducial assumptions for the case of stellar hardening since it is the most likely environmental coupling mechanism for SMBH binaries. When we increase the low-mass cutoff of systems which contribute to the characteristic strain budget this further constrains the stellar mass density. This is seen most easily in Eqn.\ (\ref{eq:fbend_stars}), where one must raise the stellar mass density to match a corresponding increase in binary mass so that the transition frequency is maintained. Furthermore, modeling the distribution of black holes masses in Eq.\ (\ref{eq:merger-density}) without the lognormal component or redshift evolution will increase the contribution of lower mass binaries to the GW strain budget, leading to smaller stellar mass density constraints than reported in Fig.\ \ref{fig:env_lims}. Varying the normalization, $a$, and exponent, $b$, of the $M-\sigma$ relation such that $a\in [7,9]$ and $b\in [4,6]$ has very little impact on the environmental constraints.

\subsubsection{Constraints on SMBH binary population eccentricity}
\label{sec:eccentricity}
It is not only the astrophysical environment of SMBH binaries that can induce a bend in the characteristic strain spectrum. Binaries with non-zero eccentricity emit GWs at a spectrum of harmonics of the orbital frequency. The cumulative effect over the entire population can lead to a depletion of the low frequency strain spectrum \citep{eenn07,s13c,rws+14,hmgt15}, and a turnover whose shape can be captured with the parametrized spectrum model employed in this paper. Hence, we can use our $f_\mathrm{turn}$ posterior from the marginalization of $f_\mathrm{bend}$ over all $\kappa$ to deduce constraints on the eccentricity of binaries at some reference orbital separation. Our approach follows from the previous astrophysics constraints, where we build populations and strain spectra which have varying eccentricities at a fixed semi-major axis of $0.01$ pc, then construct a relationship between this eccentricity and the spectral turnover. An important modeling assumption we make here is that binaries are (and always have been) driven entirely by GW emission. This factors into how we model ${\rm d}f_r/{\rm d}t_r$ and how we evolve the binary eccentricity, where we assume binaries could have eccentricities arbitrarily close to $1$ in the past. 

\begin{figure}[!h]
\begin{center}
\includegraphics[width=0.48\textwidth]{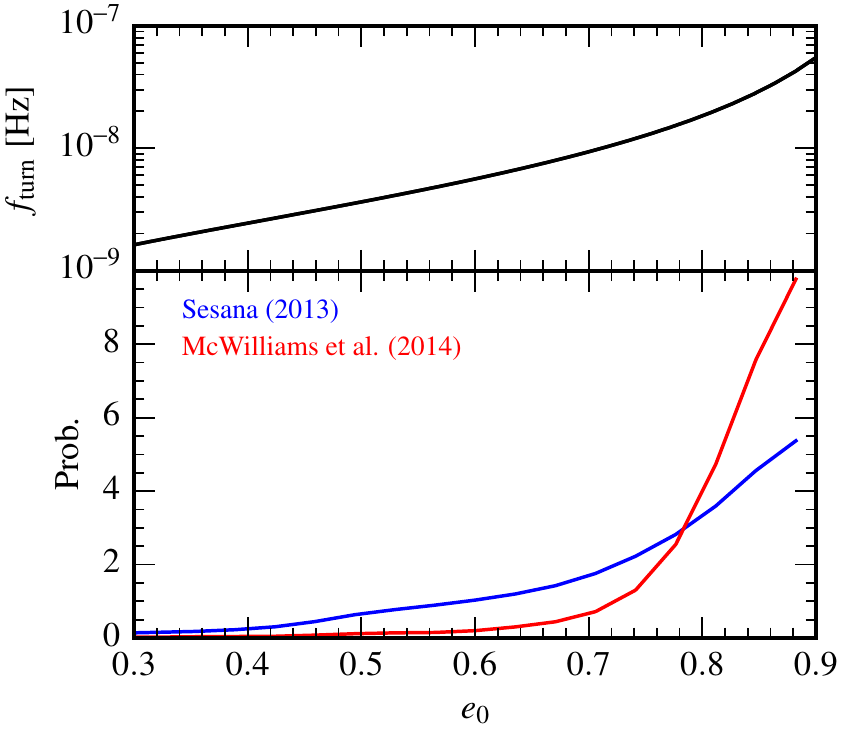}
\end{center}
\caption{Same as Figure \ref{fig:env_lims} except now we display the empirical mapping (\emph{top}) and posterior distribution (\emph{bottom}) for  the eccentricity of SMBH binaries when they had a semi-major axis of $0.01$ pc.}
\label{fig:ecc_lims} 
\end{figure}

We construct eccentric populations and corresponding strain spectra using the formalism of \citet{hmgt15}. The resulting relationship between the spectral turnover frequency and the eccentricity of all binaries at a semi-major axis of $0.01$ pc is shown in the top panel of Fig.\ \ref{fig:ecc_lims}, along with the corresponding eccentricity posteriors from each amplitude prior in the bottom panel. The high turnover frequency obtained with the \citetalias{mop14} prior leads to an eccentricity posterior distribution that largely favors $e_{0}\gtrsim 0.7$ while the \citetalias{s13} prior leads to an eccentricity posterior that is consistent with smaller eccentricities, more weakly favoring $e_{0}\gtrsim 0.5$. We emphasize that, whilst these eccentricities seem rather large, it is well established that binaries evolving in stellar environments tend to increase their eccentricity \citep{q96}. It is therefore likely that most binaries can get to $e\sim$ 0.5--0.7 along their evolution \citep[see tracks in][]{s10}. The eccentricity growth rate is generally larger for smaller binary mass ratios, and for larger initial eccentricities. The latter is indeed a crucial parameter; if, following galaxy mergers, the MBHB already has a significant ($e\gtrsim0.5$) eccentricity at the moment of formation, the subsequent evolution will almost certainly drive it to $e>0.9$. Given that the MBHB eccentricity at formation is hard to determine \citep{a03,hss02,bpb+09,amf09}, it is impossible to draw astrophysical conclusions from the constrains above. Furthermore, in reality the history of a binary's eccentricity will see phases of growth and circularization depending upon the interplay of environmental factors and GW emission \citep[e.g.,][]{s10,ks11}. This should be considered in future analyses. 



\subsection{Cosmological Model Limits}
SMBH binary mergers are not the only possible source of a GWB in the pulsar timing band. In this subsection, we use our power-law spectrum results, shown in Fig. \ref{fig:all-upper}, to constrain the fractional energy density of the Universe in relic GWs, along with stringent limits on the tension of cosmic strings.

\subsubsection{Relic GWs} 
\label{sec:relics}
Quantum fluctuations of the gravitational field in the early universe, amplified by an inflationary phase, are expected to produce a stochastic relic GW background, see e.g. \cite{g76, g77, s80, l82}. Observations of this background would provide a unique insight into the highly curved spacetime of the early universe at less than $10^{-32}$~s after the Big Bang and at energy scales of $10^{16}$~GeV,  when quantum mechanics and general relativity should reconcile, \cite{bpa+15, aab+14}. This background is expected to produce a characteristic signature in the polarization of the Cosmic Microwave Background (CMB) radiation, as well as CMB temperature anisotropies \cite{g05}. 
In the context of PTAs, the amplitude of the relic GW background is of astrophysical and cosmological interest due to the fact that it intrinsically depends on the equation of state of the early universe, $w$, and thus the Hubble constant in the inflationary stage $H_*$, as well as the tensor-to-scalar ratio $r$ , see e.g.~\cite{z11, zzy+2013}.

Specifically, the spectral index of the stochastic GWB, $\alpha$, is related to the equation of state of the early universe, $w$, via
\be
\label{eq:alpha}
\alpha = \frac{n_t}{2} -\frac{2}{3w+1}\, ,
\ee
where $n_t$ is the spectral index of the primordial power spectrum, usually set to 0. In current hot big bang models, $w=1/3$ and $n_t=0$, thus $\alpha=-1$. We therefore fix the spectral index to -1 and apply the Bayesian analysis methods described in Sec \ref{sec:bayes} to the 9-year NANOGrav dataset. Using analysis methods described in \ref{sec:bayesian-analysis}, we obtain a 95\% upper limit of on the amplitude of the relic GW background of 
\be
A^{95\%}_\mathrm{relic}=8.1\times 10^{-16} \, ,
\ee
assuming a power spectrum for the characteristic strain with $\alpha=-1$ at a reference frequency of $f={\rm yr}^{-1}$. This then constrains the 
GW energy density content per unit logarithmic frequency, divided by the critical energy density, $\rho_c$, to close the Universe:
\begin{equation}
  \label{eq:omegagw}
  \Omega_{\mathrm{gw}}(f)=\frac{1}{\rho_c}\frac{\mathrm{d} \rho_{\mathrm{gw}}}{\mathrm{d}\log f} =\frac{2\pi^2}{3H_0^2 h^2}f^2 h^2_c(f) \, ,
\end{equation}
where $f$ is the frequency, $\rho_c=8\pi/(3H_0^2)$ is the critical energy density required to close the Universe, $H_0$ is the Hubble expansion rate, and $\rho_{\mathrm{gw}}$ is the total energy density in GWs \citep{ar99, m00}. The NANOGrav limit is therefore
\be
\Omega_\mathrm{gw}(f)h^2 < 4.2 \times 10^{-10}\, ,
\ee
in a radiation-dominated universe with equation of state of $w=1/3$. This new limit is a factor of 2.9 better than the previous best limit of $\Omega_\mathrm{gw}(f)h^2 = 1.2\times 10^{-9}$ from \citetalias{ltm+15}.

Although a radiation-dominated era is usually assumed to follow inflation in the hot big bang paradigm, there is currently no evidence to show this held before Big Bang Nucleosynthesis (BBN). In fact, the existence of a reheating stage or the existence of a cosmic phase transition both violate this assumption \citep{g01, wk06,  z11}. For completeness, we now allow for other equations of state of the early universe before the BBN stage. The same analysis can be repeated assuming different equations of state for the early universe: a matter-dominated universe would have $w=0$ and therefore by Eq. \eqref{eq:alpha}, $\alpha=-2$. If the universe were instead dominated by the kinetic energy of the inflaton, then $w=1$ and $\alpha=-1/2$.

Finally, we place limits on the Hubble parameter, $H_*$ in the inflationary stage using methods developed in \citet{z11}. There, they introduced a way of translating the upper limit on a primordial GW background to a constraint on $H_*$. Using typical cosmological parameters $h=0.702$, $T_\mathrm{CMB}=0.276$~K, $\Omega_\Lambda=0.725$, $\Omega_m=0.275$, and $z_\mathrm{eq}=3454$, they obtain the following relation:

\be
\label{eq:h_star}
\log_{10}\Omega_\mathrm{gw}(k_*)=1.25-\frac{13.48}{3w+1}+2\log_{10}\left(\frac{H_*}{m_\mathrm{Pl}}\right)\, ,
\ee
where $k_*=2\pi f_*$ and is reported at a reference frequency of $f_*=1/$yr and $m_\mathrm{Pl}\equiv 1/\sqrt{G}$ is the Planck mass. Using Eq. \eqref{eq:h_star}, we can then limit $H_*$ for a fixed equation of state. For example, using the limit on $\Omega_\mathrm{gw} h^2=4.2\times 10^{-10}$ for $w=1/3$, see Table \ref{table:relicGW}, one can place a limit $H_* = 1.6 \times 10^{-2}m_\mathrm{Pl}$. Results for $w=0,1$ are in Table \ref{table:relicGW}. 

In \citetalias{ltm+15}, the limit on a primordial GW background with $w=1/3$ and $n_t=0$ is $\Omega_\mathrm{gw}(f)h^2=1.2\times 10^{-9}$, resulting in $H_*= 2.74\times10^{-2}m_\mathrm{Pl}$. One can see that the NANOGrav limit on $H_*$ is an improvement of 1.7 over the EPTA, and the most stringent limit to date.

\begin{table}
\centering
\caption{Values for $\Omega_\mathrm{gw}(f)h^2$ reported at 1/yr. Cosmological parameters used for $H_*$ are $h=0.702$, $T_\mathrm{CMB}=0.276$~K, $\Omega_\Lambda=0.725$, $\Omega_m=0.275$, and $z_\mathrm{eq}=3454$. Values of $H_*$ are in given in multiples of the Planck mass $m_\mathrm{Pl}\equiv 1/\sqrt{G}$.}
\begin{tabular}{ c  c  c  c  c }
\hline\hline
  EoS, $\omega$ & Spectral index, $\alpha$ &$A^{95\%}_\mathrm{relic}$ & $\Omega_\mathrm{gw}(f)h^2$  & $H_*/m_\mathrm{Pl}$ \\
  \hline
  0    & -2 & $1.2\times 10^{-16} $& $8.6\times10^{-12}$ & $5.4$ \\
  1/3 & -1 & $8.1\times 10^{-16}$ & $4.2\times 10^{-10}$ & $1.6 \times 10^{-2}$ \\
  1  & -1/2 &$2.0\times 10^{-15}$ & $2.5\times 10^{-9}\,$ & $8.2\times 10^{-4}$ \\
  \hline
\end{tabular}
\label{table:relicGW}
\end{table}

\subsubsection{GW Background from Cosmic (Super)strings}
\label{sec:strings}

\begin{figure*}[!ht]
\begin{center}
{\includegraphics[width=0.48\textwidth]{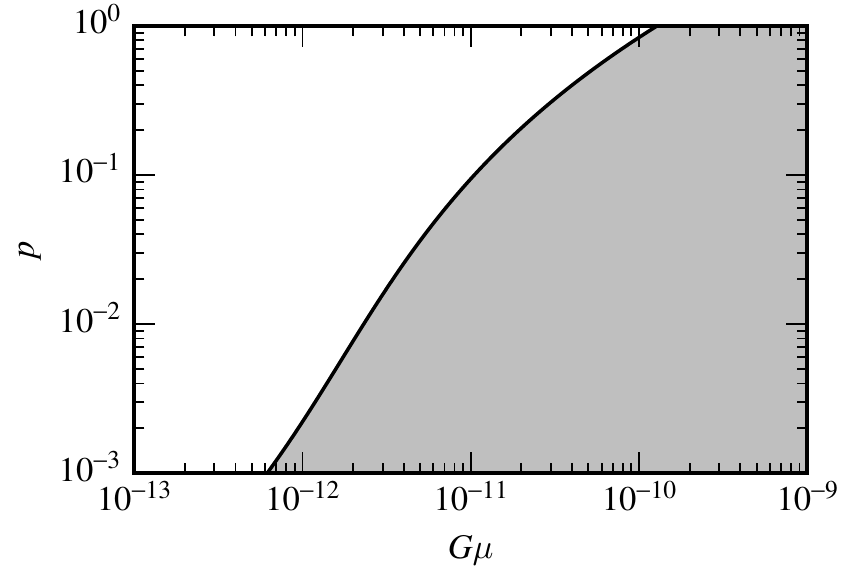}} \quad \quad
{\includegraphics[width=0.48\textwidth]{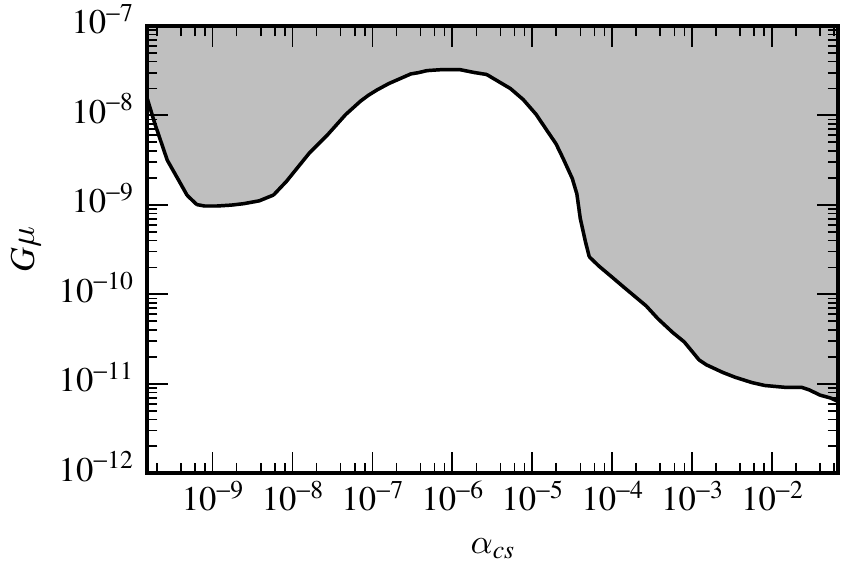}} 
\end{center}
\caption{(\textit{left}): Cosmic string constraints in terms of string tension $G\mu$ and reconnection probability $p$ using 
the results of recent cosmic string simulations described in \cite{bjo+14}.  (\textit{right}): Cosmic string tension $G\mu$ vs 
loop size parameterized by $\alpha_{cs}$  using the model described in \citetalias{ltm+15}. The shaded area is ruled out by 
our GW upper limit in both panels.}
\label{fig:strings} 
\end{figure*}

Cosmic strings are linear topological defects that can be produced in the early universe via 
phase transitions \citep{k76,v81,Vilenkin:2000jqa}. Cosmic superstrings are fundamental strings 
stretched to cosmological scales and arise in a wide range of string-theory-inspired 
cosmological scenarios \citep{st02,jst03,Copeland:2003bj}. Cosmic (super)strings produce a stochastic background of 
GWs as well as individual bursts \citep{dv01,Damour:2004kw,Siemens:2006vk,Siemens:2006yp,oms10}.

Our limits on the amplitude of the stochastic background can also be used to constrain the 
parameter space of cosmic (super)strings. Recent simulations \citep{bjo+14} have 
shown that cosmic (super)string loop densities are dominated by loops that formed at scales 
comparable to the Hubble size at the time of formation, even though only about 10\% of loops 
are formed with such large sizes.  We use the loop distributions derived by \cite{bjo+14}, specifically 
Eqs. (63), (65), and (67) of that reference with loop size $\alpha_{cs}=0.05$, together with the techniques described 
in \cite{oms10} to compute the stochastic background produced by cosmic string cusps. The 
cosmological parameters we used are taken from the Planck 2015 data \citep{planck2015}.  In 
this case the relevant parameters are the string tension $G\mu$ and the reconnection probability $p$. 
We explore this parameter space and exclude regions where the cosmic (super)string network would 
have resulted in a stochastic background amplitude larger than that ruled out by our measurements. 
The left panel of Figure \ref{fig:strings} shows the results of our analysis. On the $y$-axis we show 
the reconnection probability and on the $x$-axis the string tension. The gray shaded area shows the 
region of cosmic string parameter space that is ruled out.  Note that for $p=1$ our data only allow 
for cosmic (super)strings with tensions $G\mu < 1.3\times 10^{-10}$.  

Recently \citetalias{ltm+15} presented a comprehensive and fully general overview of 
cosmic string limits from the EPTA, and found a conservative limit on the string tension 
to be $G\mu<1.3\times 10^{-7}$. The limit is conservative in the sense that it is found 
by considering a wide range of loop sizes and taking the upper limit to be the largest possible value
 of $G\mu$ consistent with the data. The limit was 
identical to that set by the Planck Collaboration, combining Planck and high-$l$ Cosmic Microwave Background data with 
Atacama Cosmology Telescope (ACT) and the South Pole Telescope (SPT)~, cf. \cite{planck2013}. 
While the calculation in \citetalias{ltm+15} was not carried out explicitly for the \cite{bjo+14} 
simulations we can use their published limit on $\Omega_\mathrm{gw}(f)h^2=1.2\times10^{-9}$ for 
cosmic strings to place a limit of $G\mu < 8.6\times 10^{-10}$. Our limit for this model is therefore 
roughly a factor of 6.6 times more constraining than the inferred previous limit.
Using the same analysis developed by the EPTA, \citep{sbs12, sbs13}, we compute the upper limit 
on the string tension $G\mu$ as a function of loop size $\alpha_{cs}$ as shown in the right panel 
of Figure \ref{fig:strings}. Our conservative limit on cosmic string tension 
using this range of cosmic string models is $G\mu<3.3\times 10^{-8}$, a factor of 4 better 
than both the combined Planck, ACT, SPT limit and the EPTA limit.

\section{Summary and Conclusions}
\label{sec:conclusions}

This paper reports on the search for an isotropic stochastic GW background in NANOGrav's 9-year dataset. We do not find positive statistical evidence for the presence of such a signal. Following up on a series of earlier results by the three PTAs, we report new upper limits on the amplitude of backgrounds described by power-law spectra:

\begin{itemize}
\item For an astrophysical background of SMBH binaries (corresponding to a timing-residual spectral density with exponent $\gamma = 13/3$), we find a 95\% confidence limit $A_\mathrm{gw} < 1.5 \times 10^{-15}$, five times more constraining than the analogous limit for NANOGrav's 5-year dataset (DFG13).
Under the assumption of purely GW-driven evolution, leading to an unbroken $\gamma = 13/3$ power law, we compute the probability that our constraint is consistent with the \citetalias{mop14} and \citetalias{s13}/\citetalias{rws+14} theoretical predictions for $A_\mathrm{gw}$ as 0.8\% and 20\%, respectively, essentially ruling out the \citetalias{mop14} model and placing the \citetalias{s13}/\citetalias{rws+14} model in tension with our data. [Sec.\ \ref{sec:power-spectral-limits}.]
\item We verify the consistency of our limit with previously reported scaling relations between SMBH mass and galactic bulge mass, adopting fiducial estimates for galaxy merger rates and the stellar mass function. Under the assumption of circular GW-driven binaries, our limit is slightly inconsistent with the \citet{kh13} relation,
and consistent within the error margin for the \cite{mm13} relation.
[Sec.\ \ref{sec:joesarah}.]
\item We also perform an optimal-statistic (cross-correlation) analysis, and find limits that are 5.4 and 1.5 times more constraining than the analogous DFG13 and LTM15 results. The cross-correlation SNR is 0.8, indicating that there is little evidence for inter-pulsar residual correlations induced by GWs.
[Sec.\ \ref{sec:frequentist-analysis}.]
\item We extend the power-law background search to generic spectral indices, and place the most stringent limits so far on the energy density of relic GWs, $\Omega_\mathrm{gw}(f=1/\mathrm{yr})\,h^2 < 4.2 \times 10^{-10}$, for a $w = 1/3$ early-Universe equation of state. From this we obtain limits on the Hubble parameter during inflation, $H_*=1.6\times10^{-2}~m_{Pl}$. [Secs.\ \ref{sec:power-spectral-limits}, \ref{sec:relics}]

\item We place the most stringent limits to date on a GW background generated by a network of cosmic strings, $\Omega_\mathrm{gw}(f=1/\mathrm{yr})\, h^2 < 2.2 \times 10^{-10}$, which translates into a conservative upper limit on cosmic string tension $G\mu<3.3\times 10^{-8}$, using the model presented in \citetalias{ltm+15}. This is a factor of 4 better than both the combined Planck, ACT, SPT limit and the EPTA limit. Using the recent models of \cite{bjo+14} we find $G\mu<1.3\times 10^{-10}$, a factor of 6.6 times more constraining than an identical analysis using the EPTA limit. [Secs.\ \ref{sec:power-spectral-limits}, and \ref{sec:strings}.]
\end{itemize}

We further probe the interface between PTA observations and SMBH-binary population estimates by analyzing the 9-year dataset in terms of a GW background described by an inflected power law [Eq.\ \eqref{eq:turnover}]. We derive joint posteriors for the spectral parameters (the amplitude $A_\mathrm{gw}$, the inflection frequency $f_\mathrm{bend}$, and the shape exponent $\gamma$) assuming $A_\mathrm{gw}$ priors from \citetalias{mop14} and \citetalias{s13}/\citetalias{rws+14} (see Fig.\ \ref{fig:bayesogram}).
For both priors (but more so for \citetalias{mop14}), the inflected power-law model is preferred to an unbroken power law. The $f_\mathrm{bend}$ posterior can be used to infer astrophysical information about the effects that may reduce GW power at lower frequencies, such as the initial eccentricity of SMBH binaries, and the environmental influences of stars and gas in galactic nuclei. To summarize:
\begin{itemize}
\item We find that the data prefer an inflected spectrum over a pure power law, with Bayes factors $\sim$ 22 and 2.2 for the \citetalias{mop14} and \citetalias{s13} amplitude priors, respectively. The same analysis, run on the 5-year DFG13 dataset, provides little to no information about the shape of the GWB spectrum.
[Sec.\ \ref{sec:power-spectral-limits}.]
\item Under several simplifying assumptions, we map the posterior distribution of $f_\mathrm{bend}$ into posterior distributions for the nuclear stellar mass density $\rho$ (which modulates the strength of binary frequency evolution by stellar scattering, cf.\ Sec.\ \ref{sec:environs}), the SMBH mass accretion rate from circumbinary disks $\dot{M}_1$ (which can be linked to binary frequency evolution by interactions with gas, cf.\ Sec.\ \ref{sec:environs}), and the initial value of SMBH-binary orbital eccentricity $e_0$ (which distributes GW power to higher frequency harmonics, cf.\ Sec.\ \ref{sec:eccentricity}).
%
%
\end{itemize}

For the last decade (and longer), the three PTA consortia have been engaged in an accelerating race toward higher GW sensitivities---the analysis presented in this paper represents the latest stage of the race, but not the last. While our investigation cannot claim the ultimate prize, a positive detection, it is the first to use information about the amplitude \emph{and} shape of GW background to make concrete (if limited) astrophysical statements about the dynamics and environments of SMBH binaries. The era of GW astrophysics is truly upon us.

\acknowledgements


{\it Author contributions.} 
An alphabetical-order author list was used for this paper in recognition
of the fact that a large, decade-timescale project such as NANOGrav is
necessarily the result of the work of many people. All
authors contributed to the activities of the NANOGrav collaboration
leading to the work presented here, and reviewed the manuscript, text, and figures
prior to the paper's submission. Additional specific contributions
to this paper are as follows. 
ZA, KC, PBD, TD, RDF, EF, MEG, GJ, MJ, MTL, LL, MAM, DJN, TTP, SMR, IHS,
KS, JKS, and WWZ made observations for this project and developed timing
models.
JAE coordinated the writing of the paper.
JAE, RvH, and CMFM led this search by directly running the analysis pipelines.
JAE, RvH, LS, NJC, SRT, and MV designed and tuned the statistical analysis.
SRT, CMFM, STM, AS, JS, SB-S, XS, SAS, LS, MV, NJC, and JAE developed the interpretation of astrophysical results. 
CMFM developed and interpreted the relic GW results. 
XS, SAS, CMFM developed and interpreted the cosmic strings results. 
JAE, SRT, CMFM, JS, SB-S, MV, RvH, XS, STM, and AS wrote the paper, collected the bibliography, prepared figures and tables.


\emph{Acknowledgments.} The work of ZA, AB, SB-S, SJC, SC, BC, NJC, JMC, PBD, TD, JAE, NG-D, FJ, GJ, MTL, TJWL, LL, ANL, DRL, JL, RSL, DRM, MAM, STM, DJN, NP, TTP, SMR, XS, DRS, KS, JS, MV, RvH and YW was partially supported through the National Science Foundation (NSF) PIRE program award number 0968296. NANOGrav research at UBC is supported by an NSERC Discovery Grant and Discovery Accelerator Supplement and by the Canadian Institute for Advanced Research. D.R.M. acknowledges partial support through the New York Space Grant Consortium. JAE and RvH acknowledge support by NASA through Einstein Fellowship grants PF4-150120 and PF3-140116, respectively. MV acknowledges support from the JPL RTD program. Portions of this research were carried out at the Jet Propulsion Laboratory, California Institute of Technology, under a contract with the National Aeronautics and Space Administration. CMFM. was supported by a Marie Curie International Outgoing Fellowship within the 7th European Community Framework Programme. SRT was supported by appointment to the NASA Postdoctoral Program at the Jet Propulsion Laboratory, administered by Oak Ridge Associated Universities through a contract with NASA.
SAS. acknowledges funding from an NWO Vidi fellowship (PI: JTW Hessels). AS is supported by a University Research Fellowship of the Royal Society.
This work was supported in part by National Science Foundation Grant No. PHYS-1066293 and by the hospitality of the Aspen Center for Physics.
This research was performed in part using the Zwicky computer cluster at Caltech supported by NSF under MRI-R2 award no. PHY-0960291 and by the Sherman Fairchild Foundation. A majority of the computational work was performed on the Nemo cluster at UWM supported by NSF grant No. 0923409. Parts of the analysis in this work were carried out on the Nimrod cluster made available by S.M.R. Data for this project were collected using the facilities of the National Radio Astronomy Observatory and the Arecibo Observatory. The National Radio Astronomy Observatory is a facility of the NSF operated under cooperative agreement by Associated Universities, Inc. The Arecibo Observatory is operated by SRI International under a cooperative agreement with the NSF (AST-1100968), and in alliance with Ana G. M\'endez-Universidad Metropolitana and the Universities Space Research Association.

\bibliographystyle{apj}
\bibliography{bib,apjjabb}

\end{document}